\documentclass[authoryear,final,5p,times,twocolumn]{elsarticle}  
\usepackage{dcolumn}   
\usepackage{amsmath}   
\usepackage{amssymb}
\usepackage{tabularx}
\usepackage{hyperref}

\journal{Journal of Theoretical Biology}

\begin{document}
\begin{frontmatter}


\author[physics]{Sam F. Greenbury} 
\author[physics]{Iain G. Johnston}
\author[physics,caltech]{Matthew A. Smith}
\author[chemistry]{Jonathan P. K. Doye}
\author[physics]{Ard A. Louis}
\address[physics]{Rudolf Peierls Centre for Theoretical Physics, University of Oxford, 1 Keble Road,  Oxford, OX1 3NP, UK}
\address[caltech]{Division of Chemistry and Chemical Engineering 210-41, California Institute of Technology, Pasadena, CA 91125}
\address[chemistry]{Physical and Theoretical Chemistry Laboratory, Department of Chemistry, University of Oxford,   South Parks Road, Oxford, OX1 3QZ, UK}

\title{The effect of scale-free topology on the robustness and evolvability of genetic regulatory networks}

\begin{abstract}  
  We investigate how scale-free (SF) and Erd\H os-R\'enyi (ER) topologies  affect the interplay between evolvability and robustness of model gene regulatory networks with Boolean threshold dynamics.
In agreement with~\citet{cluzel_nat} we find that networks with SF$_{\rm{in}}$ topologies,  that is SF topology for incoming nodes and ER topology for outgoing nodes, are significantly more evolvable towards specific oscillatory targets than networks with ER topology for both incoming and outgoing nodes.   Similar results are found for networks with SF$_{\rm{both}}$ and SF$_{\rm{out}}$ topologies.
The functionality of the SF$_{\rm{out}}$ topology, which most closely resembles the structure of biological gene networks~\citep{babu2004}, is compared to the ER topology in further detail through an extension to multiple target outputs, with either an oscillatory or a non-oscillatory nature.
For multiple oscillatory targets of the same length, the differences between SF$_{\rm{out}}$ and ER networks are enhanced, but for non-oscillatory targets both types of networks show fairly similar evolvability.  We find that SF networks generate oscillations much more easily than ER networks do, and this may explain why SF networks are more evolvable than ER networks are for oscillatory phenotypes.
 In spite of their greater evolvability, we find that networks with SF$_{\rm{out}}$ topologies are also more robust to mutations than ER networks.  Furthermore, the  SF$_{\rm{out}}$ topologies are  more robust to changes in initial conditions (environmental robustness).
      For both topologies, we find that once a population of networks has reached the target state, further neutral evolution can lead to an increase in both the mutational robustness and the environmental robustness to changes in initial conditions.  
\end{abstract}


\end{frontmatter}

\section{Introduction}\label{intro}
\subsection{Robustness \& evolvability in genetic regulatory networks}
Biological systems require their phenotype to be robust to a variety of perturbations.  They must be {\em mutationally robust} to  minimize the possibility of a permanent deleterious mutation in the genome.  They must be {\em environmentally robust} to mitigate the effects of temporary macroscopic environmental variation, or stochastic noise in microscopic biochemical processes. By possessing a robust phenotype, functions required for survival can be performed consistently and descendants can expect a similarly reliable genotype~\citep{wag_book}.


At the same time, environments can change permanently and populations must adapt by producing heritable phenotypic variation.  This ability to generate  phenotypic innovation from genetic changes under natural selection is often called evolvability, and is observed throughout the biological world~\citep{dawkins, kirschner,wag_book}.  It may itself be an evolvable trait~\citep{earl_deem}.    The relationship between environmental robustness and evolvability is very complex~\citep{West-Eberhard2003,kirschner2005plausibility}.
On the other hand,   it would seem at first sight that evolvability and mutational robustness  are straightforwardly antagonistic:  the more easily mutations change the properties of an organism,  the less mutationally robust it should be.  

However, this simple   picture can be challenged on numerous fronts~\citep{bloom2007evolution,bloom2007neutral,wag_RS,wag_nat_pers,Daniels2008,Draghi2010}. It is important to take into account the fact that evolution acts on populations and not on individuals.  Furthermore, the structure of the evolutionary landscape, the population size $N_{pop}$, the mutation rate $\mu$ and the type of phenotypic variation required are all factors that can influence the relationship between mutational robustness and evolvability.

 For example Wagner, using RNA secondary structure as a model~\citep{wag_RS}, demonstrated the importance of  distinguishing between genotype (sequence) robustness and evolvability, which do share an antagonistic relationship, and phenotype (structure) robustness and evolvability which, by contrast, share a beneficial relationship.    For a single sequence (genotype), mutational robustness is simply the fraction of mutations that alter the phenotype. However, a single RNA phenotype (a particular secondary structure)  may be the minimal free energy structure for a large number of different genotypes.  Such a set of genotypes is called a neutral space, and during evolution, a population can spread on the neutral space to generate a large amount of genetic diversity.   The larger the neutral space, the more likely that random mutations will generate another member of the neutral space, so the larger the mutational robustness.  It can be demonstrated~\citep{wag_RS} that for RNA,   larger neutral spaces  have a larger number of  different phenotypes accessible with a single mutation. 
 Thus a population that can generate greater genetic diversity through non-deleterious mutations across a larger neutral space, also has more evolutionary innovation available, making such phenotypes more evolvable.     Furthermore, for RNA,  the utility of the neutral space with respect to robustness and evolvability is shown to be independent of the evolutionary dynamical regime (being of importance in both the $N_{pop} \mu \gg 1$ and $N_{pop} \mu\ll1$ cases, where $N_{pop}$ is population size and $\mu$ is mutation rate).
 
   These arguments depend on the structure of the neutral space~\citep{wag_nat_pers}.  If it is made up of many disconnected pockets, then a population may not be able to  explore  the entire neutral space.  If on the other hand all the neutral neighbours are clustered together tightly in one part of the space, then they may not have a large diversity of other phenotypes within easy reach.     
Models of RNA secondary structure have the advantage that they can be relatively easily analyzed by computational methods~\citep{wag_book}.       Thus detailed questions about the topology of neutral spaces and the effects of changing mutation rates and population sizes can be analyzed in some detail.   
  For other systems, however,  the relationship between mutational robustness, neutral spaces and evolvability may be more complex~\citep{Draghi2010}.

In this paper we focus on gene regulatory networks (GRNs), another set of biological systems that can be analyzed computationally.     There has been an enormous amount of interest in modeling GRNs because they play a central role in biological functionality~\citep{alon_book}.  Moreover, it is increasingly being recognized that altering gene expression patterns is a highly effective evolutionary mechanism.  Important phenotypic changes are often achievable through very small numbers of mutations to regulatory regions~\citep{carroll}.   For these reasons, the  mutational robustness, environmental robustness and evolvability of GRNs have received much theoretical attention.  For example, the reliability under perturbed system configurations  (environmental robustness) has been investigated in~\citep{aldana2003, klemm_pnas} whilst the evolution of this type of robustness is considered in~\citet{braunewell_jtbio}. Robustness to mutation has also been shown to be an evolvable property, correlating with robustness to initial gene expression states~\citep{ciliberti_plos}. Furthermore, there is evidence suggesting that neutral spaces can enhance evolvability of expression patterns~\citep{ciliberti_pnas}. The effect of varying environments has been demonstrated to increase evolvability~\citep{crombach}, whilst the role of overall network topology has also been considered~\citep{aldana2003, cluzel_nat, aldana2007}.

\subsection{Modelling GRNs}
Realistically modelling large complex interacting  GRNs is a difficult task.  Even a modest sized network can have an enormous number of parameters depending on which details of the fundamental processes involved in transcription and translation are taken into account~\citep{alon_book}.   
In practice, therefore, many different approaches to simplify the dynamics of GRNs can be found in the literature, each with their strengths and limitations~\citep{Polynikis2009a}.  


A particularly simplified model, the Boolean network, was proposed by~\citet{kauffman1969}. The generic Boolean model makes two important assumptions. First, the gene states are the variables under time evolution, as opposed to the concentrations of gene products. These states are further assumed to be either active or inactive, i.e. either ``on'' or ``off''. This approximation is based upon evidence that the changes in gene state are cooperative transitions, allowing the approximation of sigmoidal functions as step-wise ones~\citep{kauff_order}. The second assumption is synchronous updating of gene states throughout the network in discrete time steps through physically reasonable rules. In the original model, these rules were random Boolean functions of all incoming nodes, whose output determined a given node's state in the next time step. Other update rules, based upon neural threshold network models~\citep{hopfield}, have also been applied in Boolean threshold networks~\citep{wag_1994}. The Boolean model has been extensively used in theoretical biology and, despite its coarse-grained nature, has been successfully applied, for example,  to the yeast cell cycle network~\citep{li_yeast}, the segment polarity gene network of \emph{D. melanogaster}~\citep{seg_pol} and the floral cell fate of \emph{A. thaliana}~\citep{Espinosa-Soto2004}.

An important feature of Boolean network models is the manifestation of attractors. Following a certain amount of time evolution of the system, all initial configurations converge on a very small subset of possible configurations -- the attractors of the network, also referred to as the network's \emph{attractor landscape}~\citep{kauff_order}. Boolean network models have been observed to have ordered, critical and chaotic phases correlating with an increase in average connectivity~\citep{kauff_order}. In the ordered phase, small perturbations in an initial configuration tend not to spread, with the same attractor reached as in the unperturbed case~\citep{aldana2003}. However, in the chaotic phase these small perturbations spread, possibly resulting in the network arriving at a different attractor. The critical phase lies between the ordered and chaotic phases.

It has been hypothesised that the \emph{attractor landscape} represents the possible ``cell fates'' given a cell's regulatory interactions. Different sets of initial configurations induce the production of different attractors, regarded as different cell types~\citep{kauff_order}. Recently, some evidence has been produced supporting this theory~\citep{Espinosa-Soto2004,att_ev}.  Within this interpretation, there is a diverse range of possibilities, from ``frozen'' attractors of unit length, as required in developmental gene expression patterns, to ones including longer oscillatory periods capable of more complex function, for example, the changes in gene expression over the course of the cell cycle~\citep{li_yeast}.


\subsection{The role of topology}
Within the Boolean approximation that coarse-grains out many biochemical parameters, a GRN can be viewed as a directed graph, with genes as nodes and gene interactions as edges. The number of incoming and outgoing connections for a given node --- its \emph{incoming/outgoing degree} --- can be described statistically by distributions that characterize the  overall structures or topologies~\citep{barabasi_nat_rev}. 
If a set of nodes is randomly assigned connections, the resulting degrees are binomially distributed, forming Erd\H os-R\' enyi graphs~\citep{ER}. Another class is the scale-free (SF) topology, where the node degrees are power-law distributed. Given that the incoming and outgoing degrees are distinct, these two distributions can be combined to produce four different topologies, which are defined in Table \ref{top}. The probability of any SF topology occuring in a randomly constructed network decreases rapidly as the number of nodes becomes larger than the average connectivity~\citep{cluzel_nat}.

\begin{center}
\begin{table}[t]
\caption{The topologies investigated are defined with respect to combinations of incoming and outgoing node degree distributions. When average connectivity is small compared with the number of nodes, the binomial distribution may be approximated as a Poisson distribution.}\label{top}
\vspace{0.05in}
\begin{tabularx}{\linewidth}[b]{ l | X X}
Topology & Incoming degree & Outgoing degree \\
\hline
ER&Binomial/Poisson&Binomial/Poisson\\
SF$_{\rm{out}}$&Binomial/Poisson&Power-law\\
SF$_{\rm{in}}$&Power-law&Binomial/Poisson\\
SF$_{\rm{both}}$&Power-law&Power-law\\
\end{tabularx}
\end{table}
\end{center}

Since the observation of the Internet's SF topology by~\citep{barabasi1}, SF networks have been identified in many other fields, ranging from citation networks~\citep{barabasi_pa} to biological networks, including metabolic networks~\citep{jeong2000} and GRNs~\citep{babu2004}. Given the improbability of large SF topologies occurring randomly, this ubiquity has been suggested to be a consequence of common generating mechanisms~\citep{barabasi_pa}. One possible mechanism is preferential attachment, whereby highly connected nodes have a greater probability of receiving a new connection --- ``the rich get richer''. This could certainly explain the power law distribution of networks such as the Internet~\citep{barabasi_pa}, but may not be the only responsible mechanism~\citep{efk}. A modified view of preferential attachment has been suggested to be responsible for the power laws observed in large GRNs~\citep{barabasi_nat_rev}. A fundamental process in biological evolution and GRN growth is gene duplication~\citep{carroll}, where a gene is simply copied twice into the daughter system. In the corresponding GRN, all connections from the duplicated gene are doubled. As a highly connected gene is more likely to be connected to the randomly duplicated gene, new genes preferentially attach to highly connected genes.

The process outlined above would give rise to an SF degree distribution. However, in GRNs, it is only the outgoing degree that has been observed to possess a power-law distribution~\citep{babu2004,aldana2007}.  Either evolutionary rewiring has undone SF incoming degrees (due to the vast number of random network configurations relative to scale-free ones), or it is not gene duplication that is responsible for the generation of power-laws within GRNs. In either case, there must be another force either maintaining or generating an outgoing SF distrbution.


The effect of the SF topology  on Boolean networks has already been investigated by other authors.  For example, ~\citet{aldana2003} and~\citet{aldana_pnas2003} investigated the effect of dynamical perturbations in Boolean networks with an SF topology. In random networks, the ordered phase is achievable through all nodes having low connectivities, with fine tuning of mean degree connectivity. For SF topologies, however, the ordered phase can be attained whilst allowing the presence of highly connected nodes, or ``hubs''. 
In more recent work,~\citet{aldana2007} have also demonstrated increased robustness of the attractor landscape in SF topologies under the process of gene duplication.    In an important paper,~\citet{cluzel_nat} studied large ($N=500$ node) networks with Boolean threshold dynamics~\citep{kurten1988}.   They used standard genetic algorithms (GAs) to model evolution and found that when the selection pressure is for a single node in the network to produce a desired oscillatory target, an SF$_{\rm{in}}$ topology will on average reach its target in significantly less generations than an ER network will, suggesting that the former is more evolvable than the latter.  
The enhanced evolvability for oscillatory targets holds for a number of different average connectivities.  They argue that, in contrast to ER networks where the parameters must be fine-tuned to achieve a critical phase with its  associated enhanced evolvability,   the  SF network will promote evolution for a wide set of different parameter values.

In this study we consider the work of~\citet{cluzel_nat} further, with a particular focus on how the enhanced evolvability they find depends on the specific target (phenotype) and how it relates to both mutational and environmental robustness.  Besides the 
SF$_{\rm{in}}$ topology used in~\citet{cluzel_nat} we also investigate  the other topologies in Table~\ref{top} with a particular focus on the SF$_{\rm{out}}$ topology that  better resembles biological GRNs.
We note that in our study, as in~\citet{cluzel_nat}, the emphasis is on general results for very schematic models of GRNs rather than on networks directly extracted from particular experimental systems.  The advantage of this approach is that general trends are easier to extract, while the disadvantage is of course that any direct  connections to biological systems will need to be worked out further down the line.

We proceed as follows:
In Section~\ref{methods} we describe in some detail the methods that we use to simulate the evolutionary dynamics of the networks.  In Section~\ref{section3} we study the evolvability of  the four different topologies described in Table~\ref{top} for single oscillatory targets of length $L=10$, confirming that all types of SF topology are more evolvable than an ER topology.   We also demonstrate that SF topologies are much more likely to generate oscillatory outputs even without any evolutionary pressure.  
Section~\ref{multiple_targets} introduces two new kinds of targets.  For multiple oscillatory targets of the same length $L$,  the difference between ER and SF$_{\rm{out}}$ topologies is enhanced for increasing number of independent targets.    However, these differences are much smaller for multiple targets of different length.   The  difference in evolvability between the ER and the SF$_{\rm{out}}$ topologies is also quite small when evolving towards a target that fixes $100$ of the $500$ output nodes, suggesting that the enhanced evolvability of the SF$_{\rm{out}}$ topology depends in part on the kind of phenotype that is being selected for.
In Section~\ref{robustness}, the mutational and environmental robustness of the ER and SF$_{\rm{out}}$ topologies are studied.    Even though the SF$_{\rm{out}}$ topologies are much more evolvable than the ER topologies for oscillatory outputs, they are simultaneously more mutationally and environmentally robust.  
Under neutral evolution both the mutational robustness and the environmental robustness increase.
We discuss our main results in Section~\ref{conclusions}.  
To show that these results are not limited  to the model used in~\citet{cluzel_nat}, we demonstrate in~\ref{wag_model} that another GRN model~\citep{wag_1994} also exhibits a similar interplay between robustness, evolvability and topology.  Appendices B -- E are shorter and mainly focus on technical points left out of the main paper.

\begin{figure*}[!t]
\centering
\includegraphics[width=\textwidth]{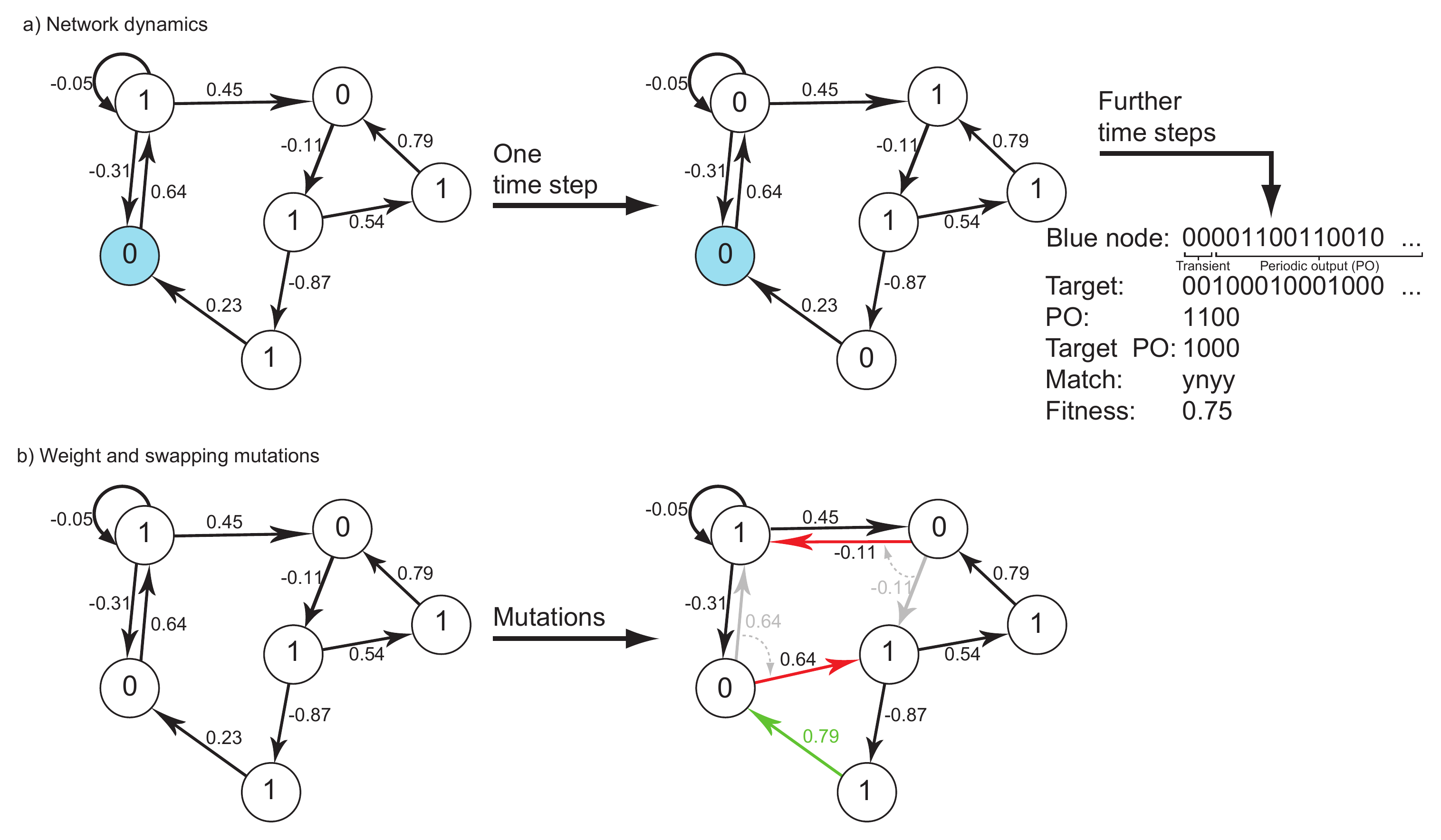}
\caption{A summary of the dynamical and evolutionary processes acting on a network. In (a) an example of the network dynamics is demonstrated. The node coloured blue is selected as the output node.  When one time step is performed, the state of the nodes will have been updated synchronously by the rules associated with Eqns. \ref{TS} and \ref{EA1}. Further times steps are then performed to produce an output bit string as shown above.  This is then compared to the target and its fitness calculated with~Eqn. \ref{hamming}. In this particular case maximal fitness is achieved. In (b) the two types of mutation are demonstrated. The red arrows represent a swapping mutation where the grey edges have been switched but preserve each node's degree distribution. The green edge, with a new weight of 0.79, represents a weight mutation from its previous value of 0.23. When a mutation occurs during an evolutionary run, one mutation type is chosen, with each type being equally probable.}\label{dynamics}
\end {figure*}

\section{Methods}\label{methods}
We model the evolutionary dynamics of a population of GRNs in several stages. Firstly, sample networks are created, generating a population of genotypes. This is followed by the simulation of network dynamics determining the attractor, with the phenotype obtained from measurements at a node  or set of nodes randomly selected at the beginning of the run. This population of networks then undergoes evolution through recurrent mutation of the genotype, followed by selection for the fittest individuals based on their phenotype, which is re-measured every generation.

\subsection{Structure}
Networks were created using the ``Configuration model'', a standard method of generating arbitrary directed networks~\citep{newman, graph_gen}. The number of nodes in the networks, $N$, was set to $500$,  whilst the average connectivity of network topologies was chosen to be $1.88$. These values provided a direct comparison with the work of~\citet{cluzel_nat}, as well as being biologically reasonable~\citep{alon_book}.

To allocate node degrees for SF topologies, numbers were drawn randomly from a power-law distribution, with the probability of degree $k$ given by
\begin{equation}\label{SFdis}
P_{SF}(k)=\frac{k^{-\gamma}}{\sum_{m=1}^{N}m^{-\gamma}}\quad1\leq k \leq N
\end{equation}
where $\gamma$ parameterises the distribution.

For the ER topology, in the case where the number of nodes is much larger than the average connectivity, the binomial distribution may be approximated as a Poisson distribution~\citep{newman}. In contrast to the SF distribution, where all nodes have at least one connection ($P_{SF}(0)=0$),  the Poisson distribution permits some nodes to have either an in or out degree of 0, or even both. Such nodes cannot both affect and be affected by the network and thus, their presence introduces a reduction in the effective size of the networks. To combat this effect, we imposed a condition on the Poisson distribution, whereby degrees of 0 are disallowed and the distribution is re-normalised accordingly. We therefore restrict ourselves to considering networks consisting of a single, large connected component. This adapted Poisson distribution for a degree $k$ is given by
\begin{equation}\label{ERdis}
P_{ER}(k)=\frac{K^ke^{-K}}{(1-e^{-K})k!}\quad1\leq k \leq N
\end{equation}
where $K$ parameterises the distribution.

The average connectivity of each distribution can be determined simply by $\langle k \rangle = \sum_{k=1}^NkP(k)$. The corresponding values of $\gamma$ and $K$ for an average connectivity of 1.88 are $\gamma=2.5$ and $K=1.431$.


\subsection{Dynamics}
A Boolean threshold model was used to model the dynamics.  At the beginning of a dynamical run, each node $\sigma_i$ is initialized  by a random process to to be either in an ``on'' state ($\sigma_i=1$) or  an ``off'' state ($\sigma_i=0$). These are the \emph{initial conditions} (\emph{ICs}) of the network. In an evolutionary run,  we either impose the same set of initial node states  at each generation (\emph{constant ICs}) or choose initial node states randomly at each generation (\emph{random ICs}). We investigate both situations in this work (in~\citet{cluzel_nat} only random ICs were used). From an initial configuration, the network dynamics are iterated over a set of discrete time steps, with the state of each node updated synchronously between time steps (see Fig. \ref{dynamics}a).

The rules for updating the state of a particular node depend on the state of the incoming nodes, and the strength of these signals. An $N\times N$ matrix of weights, $\mathbf{w}$, defines the interactions between all nodes. Connections between nodes are assigned weights in matrix $\bf{w}$ randomly from the set of real numbers on the interval $(-1,1)$. 

The state of node $i$ at the following time step is denoted $\sigma_{i}(t+1)$ and is determined by a sum over all incoming nodes
\begin{equation}\label{TS}S_i=\sum_{j=1}^{N}w_{ij}\sigma_{j}(t)\end{equation}
combined with the following threshold rule:
\begin{equation}
\label{EA1}
\sigma_i(t+1) = 
\begin{cases} 
1 &\text{if } S_i > 0 \\
\sigma_i(t) &\text{if } S_i=0 \\
0 &\text{if } S_i < 0  
\end{cases}
\end{equation}
To physically interpret the matrix of weights, consider a connection from node \emph{i} to node \emph{j}. If $w_{ij} > 0$, node $j$ promotes node $i$ whilst for $w_{ij} < 0$, node $j$ inhibits node $i$. If $w_{ij}=0$ there is no connection from \emph{j} to \emph{i}.

To prevent frozen dynamics, there is a modified rule for nodes with only a single incoming connection~\citep{cluzel_nat}.
 If node $j$ is incoming to
node $i$ the the dynamics depend purely on the sign of the weight with
\begin{equation}
\sigma_i(t+1) = 
\begin{cases} 
\sigma_j(t) &\text{if } w_{ji} > 0 \\
\neg \sigma_j(t) &\text{if } w_{ji} < 0,
\end{cases}
\end{equation}
where $\neg$ is the NOT operator. The behaviour of nodes connected by this modified rule is simple. Node states are either copied or inverted for positive and negative weights respectively.


The critical parameter values, $\gamma_c$ and $K_c$, separating the ordered and chaotic phases for each topology, have been previously calculated by~\citet{cluzel_nat} based upon the annealed approximation introduced by~\citet{derrida86}. We repeat the calculation (see \ref{ord_cha}) finding the critical value for the SF topology to be $\gamma_c=2.42$ for networks of size $N=500$. We also calculate the critical connectivity for the adapted Poisson distribution, which we determine to be $K_c=3.538$. The primary parameter values used in this study of $\gamma=2.5$ and $K=1.431$, both produce average connectivities below that needed for criticality, so that the networks are within the ordered regime.

After a run typically lasting 350 time steps, the steady state dynamics of the output node(s) in each network is determined and this sequence defines the individual's phenotype. An example of a single dynamical time step is performed on model networks in Fig. \ref{dynamics}a.


\subsection{Evolution}
To model evolution of the population towards a desired output (phenotype) we used a simple genetic algorithm.  First a population of $N_{pop}=50$ independent networks is generated.  Then, for each generation, the following cutoff selection regime is applied. Each network first produces three daughter networks, enlarging the population from $N_{pop}=50$ to $N_{pop}=200$. Each node in these daughter networks has a \emph{mutation probability} $\mu = 0.02$ of undergoing a mutation, a value previously used by~\citet{cluzel_nat}. There are two types of mutation: either a random change in the weight of any of a node's connections or an incoming connection is swapped with another incoming connection in the network, provided the new connections do not already exist (Fig. \ref{dynamics}). Both mutation types preserve the node degree distribution throughout an evolutionary run.

In single target experiments, the output of a node, randomly selected at the start of the evolutionary run, is used to calculate the network's fitness. The fitness is calculated from the minimum Hamming distance between the node's periodic output and the target output over all cyclic permutations. We define this mathematically as
\begin{equation}
\label{hamming}
F = \max_{j \in \{0, ..., \delta\} } \left( 1 - \frac{1}{L_o L_{tgt}} \sum_{i=1}^{L_o L_{tgt}} |\sigma_o(i + j) - \sigma_{tgt}(i)| \right),
\end{equation}
where $L_o$ is the output period, $L_{tgt}$ is the target period, $\sigma_o(i)$ is the $i$th bit in the output sequence and $\sigma_{tgt}(i)$ is the $i$th bit in the target sequence. The sequences are, for comparison, extended by repeating them periodically to construct sequences of equal length $L_o L_{tgt}$. The sum is then a Hamming distance of these two strings. Taking the maximum over an offset $j$ between 0 and $\delta = \mbox{max}\{L_o, L_{tgt}\}$ performs the Hamming distance calculation over all cyclic permutations of the strings. 

The Hamming distance measures the fraction of the sequence that matches the target, providing a linear measure for the fitness. Once all networks in the enlarged population have been assigned a fitness, the networks are ordered by fitness and the top 50 kept to form the population for the next generation.

\section{Evolution towards single L=10 targets}\label{section3}
\subsection{All types of SF topology are more evolvable}\label{sec:1x10}
Expanding on the work of~\citet{cluzel_nat}, we tested evolutionary performance  or evolvability of each of the four different topologies (ER, SF$_{\rm{out}}$, SF$_{\rm{in}}$ and SF). Evolvability is measured through adaptation speed towards the target output, i.e.\ the number of generations before every member of a population has reached maximum fitness.  
Evolvability has been frequently used in different contexts, sometimes creating confusion~\citep{poole,wag_book}.  In this case, however, the adaptation speed would seem the most natural measure of evolvability.

A randomly chosen node from each network was evolved towards a randomly chosen $L=10$ target for a maximum of $10\,000$ generations. This was repeated 50 times for each topology, with both constant and random ICs. When evolution is performed with random ICs, robustness to this stochasticity must evolve for an individual to maintain itself in the population. For constant ICs there is no such constraint.

We observed that occasionally an evolutionary run continues for a very large number of generations before converging to a solution, or sometimes not converging at all within the limits of our calculations. The mean of the adaptation time distribution can be dominated by these rare events.  For that reason, we use the median adaptation time, arguing that this  provides a more accurate measure of the typical adaptation time than the mean does.   We define $\tau$ to be the number of generations for all members of a population to reach maximum fitness in a single evolutionary run, whilst $\tilde{\tau}$ is the median over a batch of runs. The values of $\tilde{\tau}$ for the different topologies and ICs, are presented in Table \ref{1x10_meds}.   

Out of the three SF topologies, the SF$_{\rm{both}}$ evolves more rapidly  than the SF$_{\rm{in}}$ topology does, which in turn is followed by SF$_{\rm{out}}$ and ER. This result supports the findings of~\citet{cluzel_nat}, where the SF$_{\rm{in}}$ topology is tested against the ER topology and shown to evolve much more rapidly. The order of the topologies, with respect to adaptation speed, is unchanged between constant and random ICs, although the decrease in speed from constant to random ICs varies amongst the topologies, with SF$_{\rm{in}}$ and SF$_{\rm{both}}$ topologies showing greater relative changes than ER and SF$_{\rm{out}}$ do. This latter property gives an indication of the natural robustness to ICs of each topology, a property we explore in later sections.

\begin{center}
\begin{table}[t]
\caption{The median number of generations to reach maximal fitness, $\tilde{\tau}$, for runs evolving towards a single $L=10$ target. Results are shown for both random and constant ICs. All SF topologies outperform the ER topology. The SF$_{\rm{in}}$ topology is most significantly affected by the difference between constant and random ICs, whilst the SF$_{\rm{out}}$ is least affected.}\label{1x10_meds}
\vspace{0.05in}
\begin{tabularx}{0.95\linewidth}[b]{| l | X | X | X | X |}
\hline
ICs & ER & SF$_{\rm{out}}$ & SF$_{\rm{in}}$ & SF\\ \hline
Constant & $950$ & $450$ & $51$ & $25$ \\
Random & $3\,200$ & $680$ & $390$ & $200$ \\ \hline
Relative & $3.4$ & $1.5$ & $7.7$ & $8.0$\\
\hline
\end{tabularx}
\end{table}
\end{center}

\subsection{The SF topology exhibits more oscillations and greater synchronicity than the ER topology}\label{sf_adv}
\begin{figure}[t]
\centering
\includegraphics[width=\linewidth]{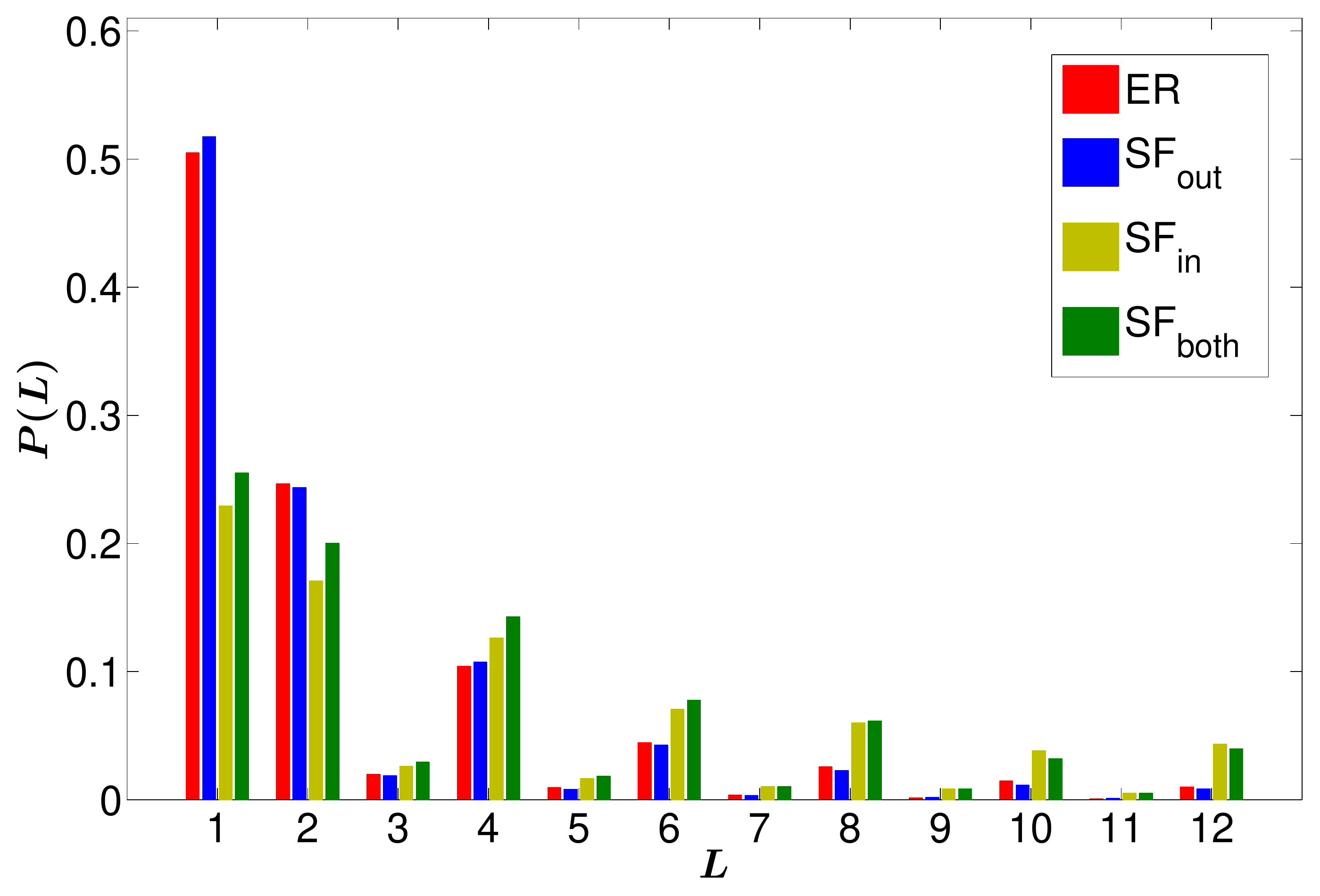}
\caption{The probability of different periods being the largest in randomly generated networks. SF$_{\rm{both}}$ and SF$_{\rm{in}}$ perform similarly, whilst SF$_{\rm{out}}$ and ER are similar too.}\label{osc_lrep}
\end{figure}

\begin{figure}[t]
\centering
\includegraphics[width=\linewidth]{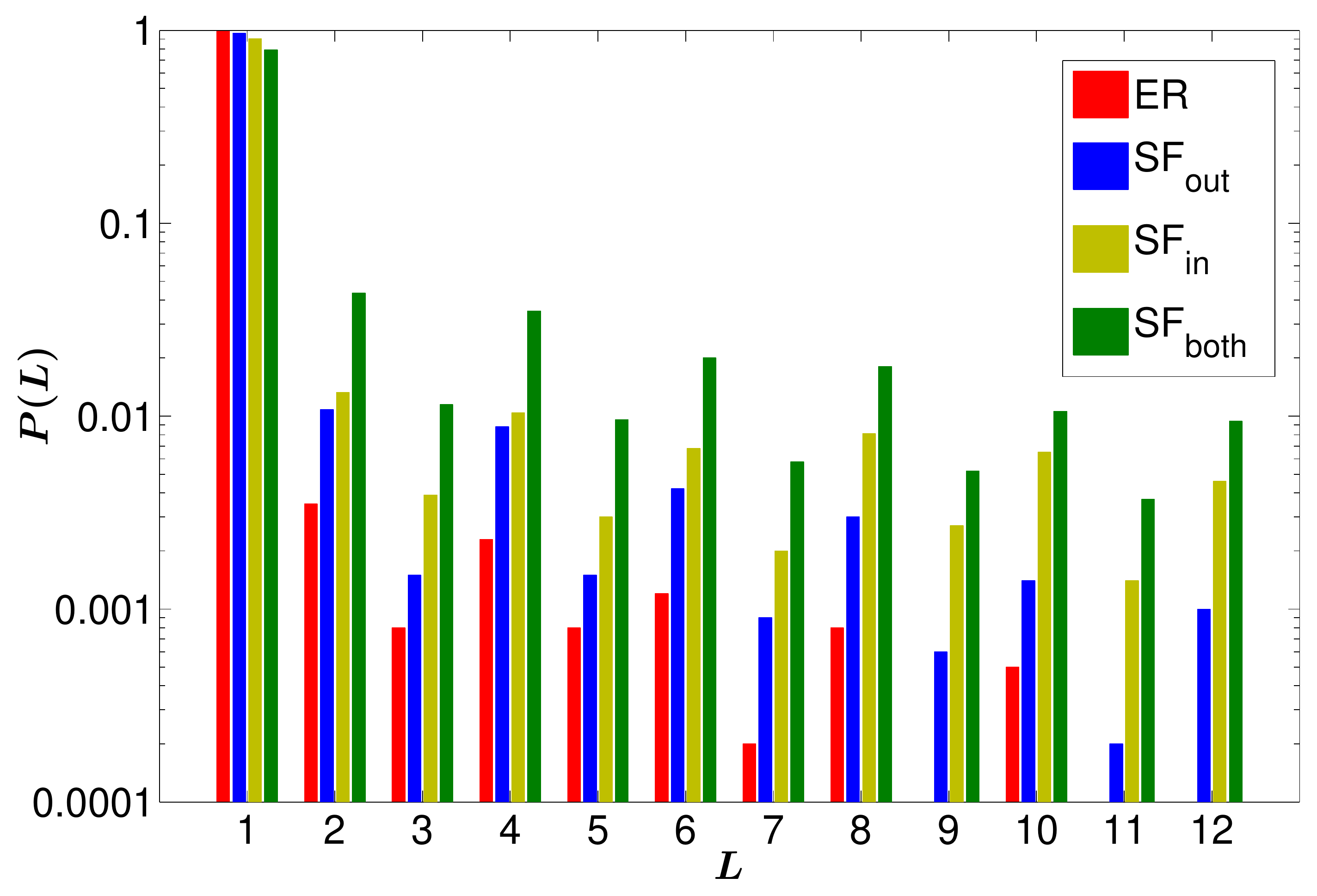}
\caption{The probability of different periods being measured at a random node, in randomly generated networks. Note the logscale. SF$_{\rm{both}}$ generates the most oscillations, followed by SF$_{\rm{in}}$, SF$_{\rm{out}}$ and ER. 
}\label{osc_rep}
\end{figure}

To investigate whether the SF topology more naturally generates oscillations than the ER topology does, we examined the dynamics of networks without evolution towards a target.
%
%
For each topology $10\,000$ networks were generated and two measurements made. For the first, the period of every node in a network was found and the largest period recorded. A histogram of these periods is presented in Fig. \ref{osc_lrep}. For the second measurement, the period of a randomly chosen individual nodes was recorded. A histogram of periods from these $10\,000$ networks is presented in Fig. \ref{osc_rep}.

In Fig. \ref{osc_lrep}, we observe  a clear divide between the SF$_{\rm{out}}$ topology and the other two SF topologies. The SF$_{\rm{out}}$ topology produces a probability distribution similar to that of the ER topology. The SF$_{\rm{both}}$ and SF$_{\rm{in}}$ topologies have similar distributions to each other. They are significantly more likely to have maximum periods at larger values  of $L$.   


Fig. \ref{osc_rep} examines the probability to find oscillations of period $L$ at a randomly chosen node. This result differs to the previous one as SF$_{\rm{both}}$ now shows more oscillations than SF$_{\rm{in}}$, whilst SF$_{\rm{out}}$ is clearly superior to ER at generating oscillations.   Again the SF$_{\rm{in}}$ and SF$_{\rm{both}}$ topologies are considerably more likely than the SF$_{\rm{out}}$ topology to have an oscillating node. 
In both Fig.~ \ref{osc_lrep} and~\ref{osc_rep} even oscillations are more easy to generate than odd ones.


\begin{figure}[t]
\centering
\includegraphics[width=\linewidth]{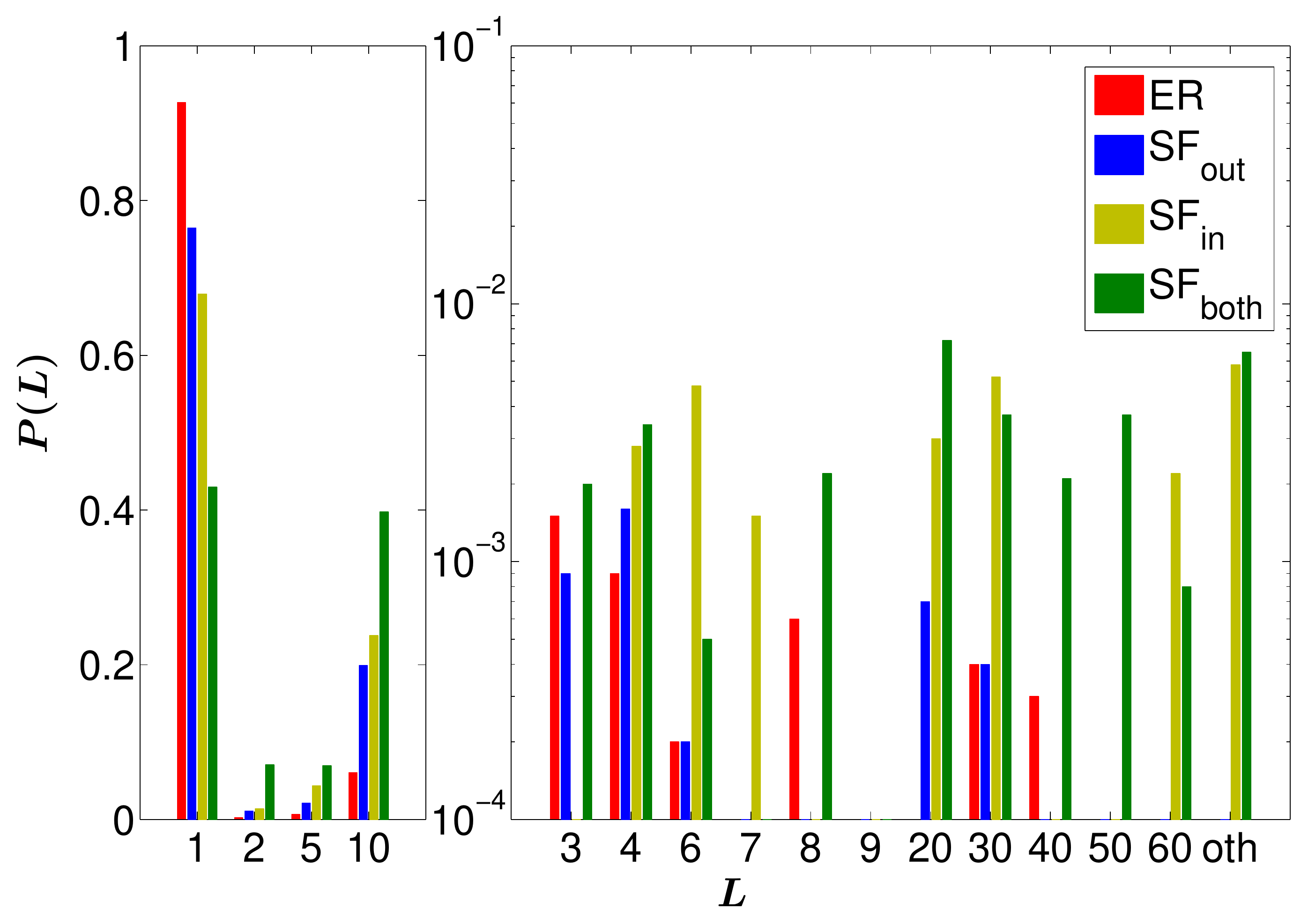}
\caption{The probability of a randomly chosen node possessing period $L$ in the first generation where maximum fitness for an $L=10$ target is attained. The probability is calculated by taking  the mean fraction of nodes with each period, taken over 50 independent evolutionary runs. The ``oth'' column refers to other periods, i.e. all periods not labelled.   This figure measures the degree of synchronicity in the networks.  Note the difference in scale on the y-axes of the right and left graphs.}\label{sync}
\end{figure}

We next measured the distribution of oscillations in the network during evolutionary runs towards $L=10$ targets.  The period of each node was measured in the first generation where all individuals in the population have a fitness of unity (defined to be the first \emph{maximally fit} generation), giving an indication of the extent to which the target node's period is spread throughout the network. This measures the \emph{synchronicity} within evolved networks.

The mean fraction of nodes oscillating with each period are presented in Fig. \ref{sync} for each of the four network topologies, averaged over 50  independent evolutionary runs. This figure demonstrates that a greater fraction of ER networks are frozen ($L=1$). All the SF topologies have an increased number of oscillatory nodes and particularly $L=10$ nodes. Periods that are factors of 10 (2 and 5) also appear to be much more prevalent than other periods. 

In general we observe that the scale-free topologies are more likely to exhibit oscillations at a randomly chosen node, have larger maximum periods and show more synchronicity during an evolutionary run.
In general the differences with the ER networks are largest for the  SF$_{\rm{in}}$ and SF$_{\rm{both}}$ topologies.   In that context it is important to recall that there is a modified update rule for single incoming nodes.  Through this rule, any oscillatory node connected to a node with a single incoming connection will always propagate the oscillation to that node. Given that $P_{SF}(1)=0.745$ compared with $P_{ER}(1)=0.450$, guaranteed propagators are much more likely in SF$_{\rm{in}}$ networks.
 The adaptive advantage generated by the SF$_{\rm{out}}$ topology must have different origins, however, as the corresponding $P(1)$ is the same as for ER networks, and hence the modified rule affects both cases similarly.

In the rest of this paper we focus mainly on comparisons of the ER with the SF$_{\rm{out}}$ distribution because the  latter most closely resembles the topology of  GRNs found in nature~\citep{babu2004,aldana2007}.


\section{Evolution towards multiple oscillatory and frozen targets}\label{multiple_targets}
In nature GRNs may employ phenotypes where multiple nodes require oscillating periods. This would be the case, for example, if a set of genes played a part in an oscillatory process within an organism. The segmentation clock~\citep{seg_clock} and circadian rhythms~\citep{circadian} rely on multiple genes working together in this way. 

To study this scenario, evolution was performed towards a target phenotype
defined by requiring the output of several randomly chosen nodes'  to have specific periods.  We define the \emph{target period set}, $\mathcal{L}$, to comprise the period of each target node. For example, for $N_{\text{tgts}}=3$ with two $L=5$ targets and one $L=10$ target, $\mathcal{L}=\{5,5,10\}$.

To determine the fitness of a network, each target is assumed to make an equal contribution. A random set of nodes are assessed against each of the randomly chosen target outputs of length specified in  $\mathcal{L}$ and each node's fractional fitness is added. As such, we define fitness of the networks with multiple targets as \begin{equation}F=\sum_{i=1}^{N_{\text{tgts}}}F_i/N_{\text{tgts}}\end{equation}
where $F_i$ is the fitness of the $i^{\text{th}}$ node in the phenotype and $N_{\text{tgts}}$ is the number of target nodes defining the phenotype. Each $F_i$ is given by Eq. \eqref{hamming}.

\subsection{The adaptation time required to achieve multiple targets of the same length increases rapidly with the number of targets}
As the ER topology evolves rather slowly towards $L=10$  targets, a smaller target length was selected. This allows a greater range of target numbers to be tested within a computationally reasonable number of generations. If the target length is too small, however, only a few different targets will be possible. For example, there are only 6 possible $L=5$ targets (see \ref{number_of_outputs}). Given these considerations, multiple $L=7$ targets were chosen as this length is easily attainable by the ER topology whilst there are 18 possible $L=7$ targets.

\begin{figure}[t]
\centering
\includegraphics[width=\linewidth]{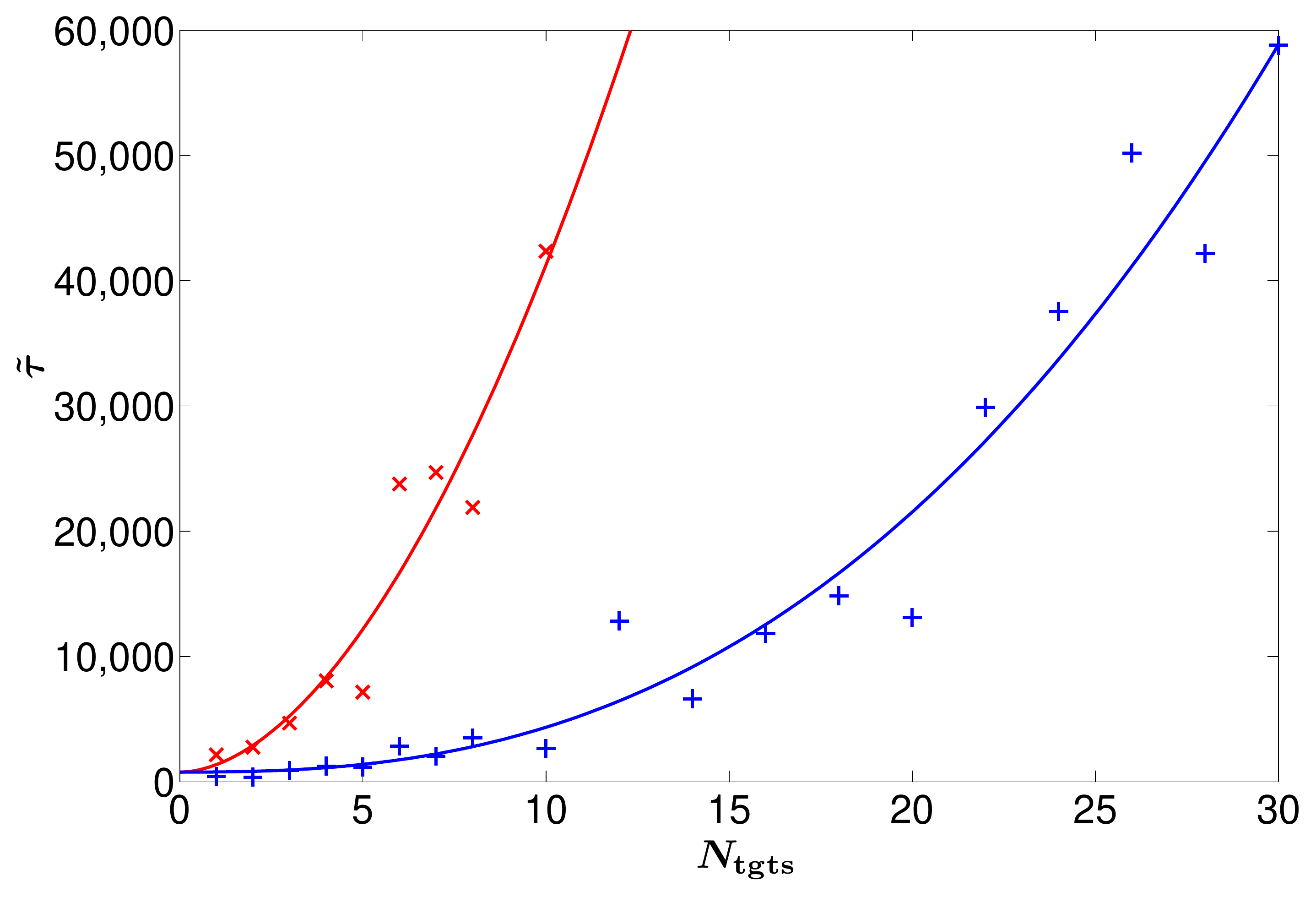}
\caption{$\tilde{\tau}$ v.s.\ the number of $L=7$ targets for the ER (red, $\times$ markers) and SF$_{\rm{out}}$ (blue, $+$ markers) topologies with corresponding  fits (the lines are guides to the eye). Both topologies experience an increase in adaptation time with an increase in the number of targets. This increase is more rapid for the ER topology, typically requiring more than $20\,000$ generations for $N_{\text{tgts}}>6$. The slower rate of increase of $\tilde{\tau}$ seen in the SF$_{\rm{out}}$ topology is likely to be due to the greater synchronicity of this topology.}\label{nx7mers}
\end{figure}

To measure speed of adaptation, we used the median number of generations until a population is maximally fit, $\tilde{\tau}$, as in Section \ref{sec:1x10}. In Fig. \ref{nx7mers}, $\tilde{\tau}$ is plotted against $N_{\text{tgts}}$. As expected, for both ER and SF$_{\rm{out}}$ topologies there is an increase in the adaptation time as $N_{\text{tgts}}$ increases. The change in adaptation time is non-linear, increasing rapidly with the number of targets. The rate of increase is much more rapid in ER networks than SF$_{\rm{out}}$ ones.  For example,  SF$_{\rm{out}}$ networks are still capable of adapting to $N_{\text{tgts}}=20$ in just over half the time ER takes to adapt for $N_{\text{tgts}}=8$.

As observed earlier in Section \ref{sf_adv}, the SF$_{\rm{out}}$ topology shows greater synchronicity for oscillatory signals than the ER topology does.    This difference may explain why  the SF$_{\rm{out}}$ topology performs so much better for multiple targets of the same length.

\subsection{The adaptation time required to acheive multiple targets of different lengths depends on the number of targets and their relative periods}

To test whether the enhanced synchronicity of the SF$_{\rm{out}}$ topology is a key reason it performs so much better for multiple targets of the same length, we also examined different sets of multiple targets 
with different lengths. All the evolutionary runs were performed with random ICs, and the results are shown in  Table~\ref{mdl_meds}.     Clearly having targets with multiple lengths greatly slows down the median adaptation time  compared to having multiple targets of the same length for both topologies. It is also interesting to compare the performance of the networks for the $\mathcal{L}=\{5,10\}$ targets to the performance for the $\mathcal{L}=\{5,8\}$ targets.   Although the former targets are slightly longer than the latter, they are easier to reproduce.   Presumably this is because $L=10$ is a multiple of $L=5$ whereas $L=8$ and $L=5$ are coprime.     We also note that the difference between the SF$_{\rm{out}}$ and ER topologies is  smaller for the coprime targets, suggesting that the enhanced synchronicity of the SF$_{\rm{out}}$ topology, observed in Fig.~\ref{sync}, plays an important role in its enhanced ability to adapt to certain types of targets (phenotypes).
 Finally, we also tested the target $\mathcal{L}=\{5,10,20\}$ but were not able to find solutions within the $40\,000$ generation cutoff we used in our simulations, showing that multiple longer targets are significantly more difficult to achieve than a larger set of shorter targets (as in Fig.~\ref{nx7mers}).

\begin{center}
\begin{table}[t]
\caption{Values of $\tilde{\tau}$ for evolutionary runs evolving under random ICs to different sets of periods. $\tilde{\tau}$ is smaller for $\mathcal{L}=\{5,10\}$ compared with $\mathcal{L}=\{5,8\}$ for both topologies.   For the $\mathcal{L}=\{5,10,20\}$  no convergence was achieved.
 }\label{mdl_meds}
\vspace{0.05in}
\begin{tabularx}{0.85\linewidth}[b]{| X | X | X |}
\hline
Period set, $\mathcal{L}$ & ER & SF$_{\rm{out}}$ \\ \hline
$\{5,10\}$& $7\,000$ & $2\,400$ \\
$\{5,8\}$ & $9\,900$ & $4\,700$ \\
$\{5,10,20\}$ &$> 40\,000$&$> 40\,000$\\
\hline
\end{tabularx}
\end{table}
\end{center}

\subsection{The evolvabilities of SF$_{\rm{out}}$ and ER networks for multiple $L=1$ targets are much more similar than for oscillatory targets}
Thus far only targets of an oscillatory nature have been considered. This is because single $L=1$ targets would be trivial to evolve; nodes of $L=1$ are by far the most common in all attractors. With the extension to multiple target nodes, however, evolving to a large number of specific $L=1$ targets is a non-trivial task. Biologically, a set of ``frozen'' targets could correspond to a particular  constant gene expression pattern over some length of time, for example in an organisms' development.

As was done earlier, networks of each topology were evolved over $10\,000$ generations in 50 independent runs. 
Out of the $N=500$ nodes, 100 were randomly chosen and required to possess a specific $L=1$ target. We will refer to this target set as $\mathcal{L} = \{100 \times 1\}$. The values of $\tilde{\tau}$ with random and constant ICs are displayed in Table \ref{100x1_meds}. The SF$_{\rm{out}}$ topology maintains its advantage over the ER topology but the difference in evolutionary adaptation times are vastly reduced. 
The effect of random ICs does not slow evolution down that much, and there is a similar relative decrease for both topologies.

\begin{table}[t]
\caption{ Median adaptation time $\tilde{\tau}$ for runs evolving towards $\mathcal{L}=\{100 \times 1\}$ targets. The performance of the topologies is much more similar than for evolution towards oscillatory targets, although the SF$_{\rm{out}}$ topology retains a slight advanatage. 
}\label{100x1_meds}
\vspace{0.05in}
\begin{centering}
\begin{tabularx}{0.9\linewidth}[t]{| X | X | X | }
\hline
ICs & ER & SF$_{\rm{out}}$ \\ \hline
Constant & $255$ & $193$ \\
Random & $402$ & $278$ \\ \hline
Relative & $1.58$ & $1.44$ \\
\hline
\end{tabularx}
\end{centering}
\end{table}

\section{Robustness in single \& multiple target experiments}\label{robustness}

In the previous section we have investigated the evolvability of ER and various SF topologies towards oscillatory and frozen targets.    In this section we study the mutational and environmental robustness of these networks under similar evolutionary situations.   We particularly focus on the change in robustness under neutral evolution, after the population has achieved maximal fitness.



\subsection{Mutational robustness is higher in SF topologies and increases  under neutral mutation}\label{r_m}

\begin{figure}[t]
\centering
\includegraphics[width=\linewidth]{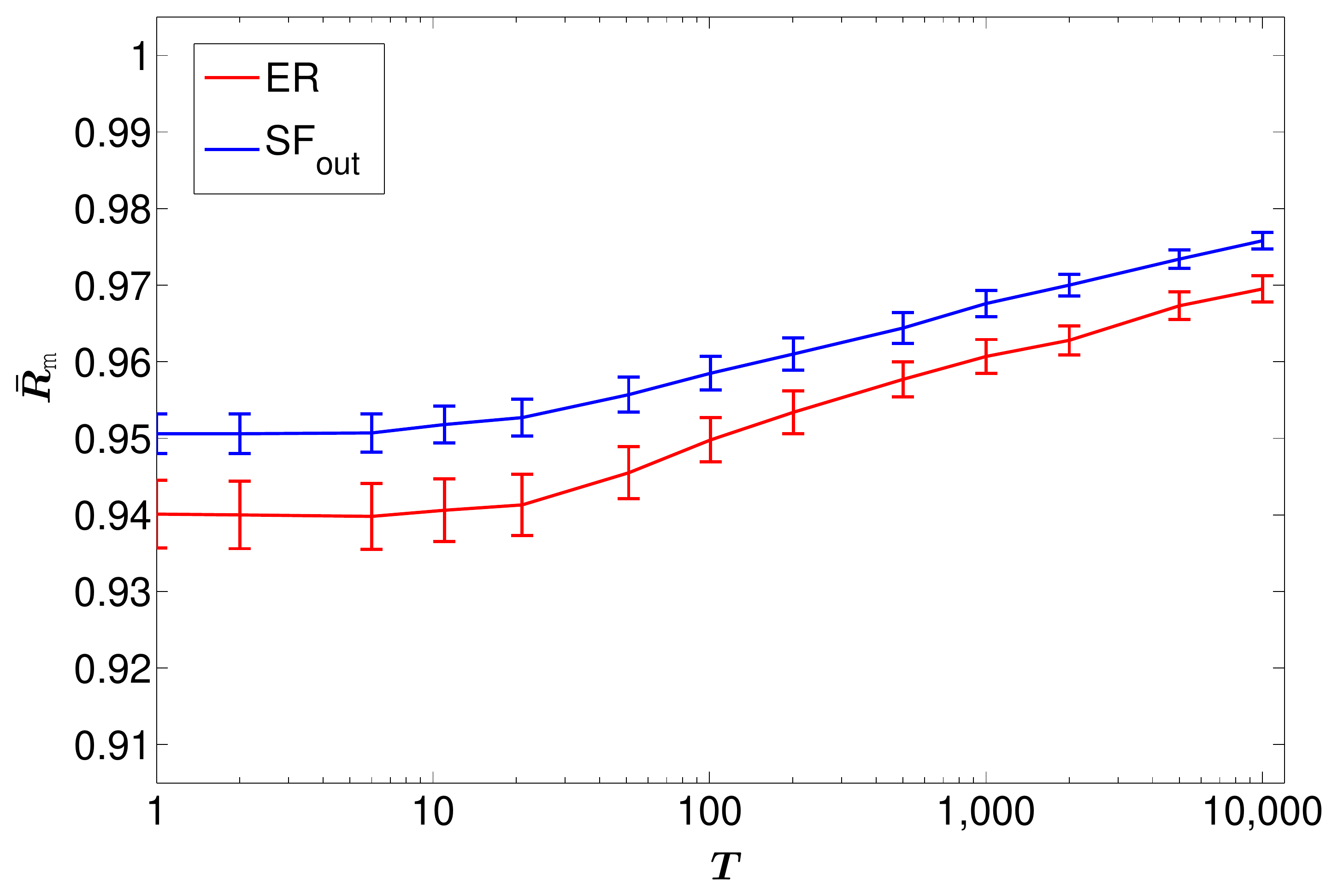}
\caption{Mean mutational robustness $\bar{R}_m$, against generation $T$ (where $T=1$ is the first maximally fit generation) for $\mathcal{L}=\{10\}$. The errorbars represent the standard error on the mean. Robustness is still increasing after $10\,000$ generations for both topologies. The SF$_{\rm{out}}$ topology is more robust from the outset and remains so throughout. Both topologies are able to evolve greater robustness. Noting the $y$-axis range, the change in robustness is small but statistically significant for both topologies.}\label{r_m_1x10}
\end{figure}

To examine the mutational robustness of large networks with a large range of weights, a sampling approach must be taken as the number of possible network genotypes in a 1-mutation neighbourhood is astronomical. To sample these genotypes, each node in a network was tested with 6 sample mutations of each mutation type in Fig. \ref{dynamics}. We employ constant ICs throughout, ensuring that IC robustness effects do not play a role.
An individual's mutational robustness, $R_m^{ind}$, is defined as the proportion of sample mutations that do not result in a network's phenotype changing. The robustness testing was performed on every member of a population during an evolutionary run at the generation where maximum fitness is attained ($T=1$), and then at intervals ($T = 2, 5, 10, 20, 50, 100, 200, 500, 1\,000, 2\,000, 5\,000, 10\,000 $) afterwards. At each interval, $R_m^{ind}$ is averaged over the whole population producing an average population mutational robustness, $R_m$. By averaging over the population, the  statistical error is reduced.  We show in~\ref{diversity}  that the diversity of genotypes is maintained during neutral evolution. 

Mutational robustness was measured for $\mathcal{L}=\{10\}$ and $\mathcal{L}=\{100 \times 1\}$ targets, although we only show results for the former target  in Fig. \ref{r_m_1x10}.
The main and perhaps unexpected result for both types of target is that throughout these evolutionary periods the SF$_{\rm{out}}$ topology maintains a greater robustness to random mutation than the ER topology, even though it is simultaneously more evolvable.  We also note that $R_m$ is  fairly high for both topologies, suggesting that the vast majority of mutations don't affect the phenotype. The most likely reason for this is that a large number of nodes simply do not participate in the core network that drives the output of the target node.   Although it is not shown, the value of $R_m$ is also fairly  high for the $\mathcal{L}=\{100 \times 1\}$  targets ($R_m > 0.8$) and moreover the SF$_{\rm{out}}$ topology outperforms the ER topology here as well.  Again, since $20\%$ of the nodes are fixed by the target output, this suggests that a large number of the nodes do not participate in the output that determines the phenotype of these systems.


The second effect this figure clearly shows is that the mutational robustness increases with evolutionary time.   It was shown in~\citet{van_nim} that under neutral mutation, the mutational robustness of a population can increase.   One mechanism for this phenomenon is that those genomes that are more mutationally robust are more likely to have offspring that share the same phenotype.   Under continued selection then, such genomes are more likely to survive than genomes with low mutational robustness.  These authors also showed that for selection to optimize mutational robustness in  neutrally evolving systems, the condition  $\mu N_{pop} \gg 1$  must be satisfied. Here the mutation rate per node is set to $\mu_{node}=0.02$. For an $N=500$ node network, this gives an average of $\mu = \mu_{node} N=10$  mutations per mutated offspring. With a fraction $m=0.75$ mutated offspring in the extended population before the cut off is imposed,  in each generation there are $\mu\times m\times N_{pop}=\langle\mu\rangle N_{pop}=10\times0.75\times50=375$ mutations in a future generation's genepool.  
Given that the condition $\mu N_{pop} \gg 1$  is satisfied, along with the genetic diversity observed in the populations (see \ref{diversity}), further optimisation of mutational robustness is certainly expected. However,  selection may not be the only cause of the observed effects. It may be possible for the increase to be obtained purely through entropic effects -- for example, if many more genotypes are available at higher mutational robustness, the population will discover such possibilities simply through diffusing over the neutral space. Working out  the exact reasons of this optimisation is thus a subtle question, and we don't think that our results unambiguously fix the cause of the increase in $R_m$.

\subsection{Environmental robustness increases markedly with time after maximal fitness is reached for random initial conditions, and is higher in SF topologies}\label{r_ic}
\begin{figure}[t]
\centering
\includegraphics[width=\linewidth]{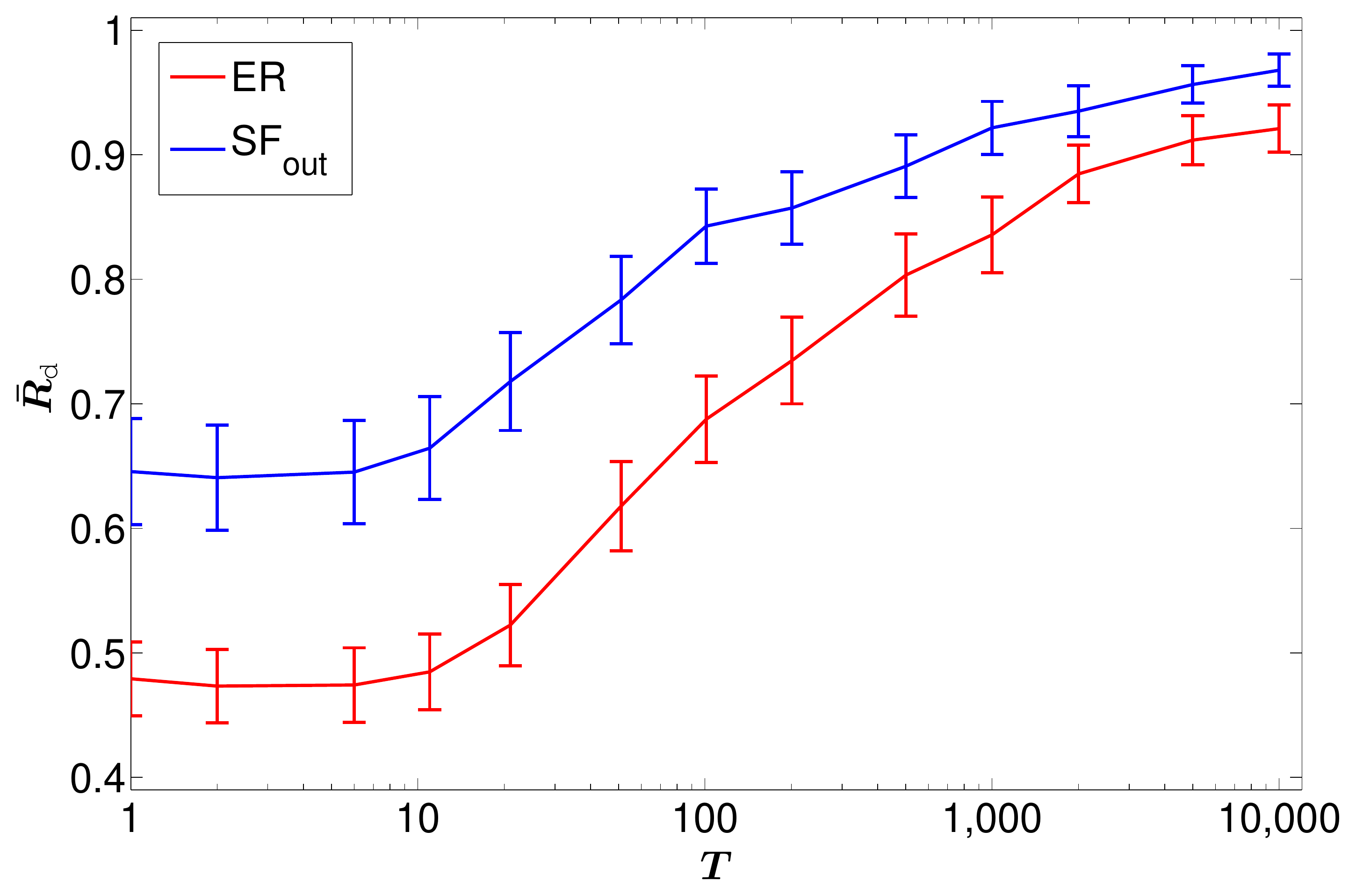}
\caption{Mean environmental robustness $\bar{R}_d$, against $T$ with random ICs for $\mathcal{L}=\{10\}$. SF$_{\rm{out}}$ is initially much more robust, although the gap is reduced over the course of neutral evolution. The increase in $\bar{R}_d$ is dramatic for both topologies, indicating a strong optimisation.}\label{r_ic0_1x10}
\end{figure}

GRN oscillators \emph{in vivo} are often required to function under a variety of environmental conditions. By possessing environmental robustness, function can be maintained under such stochasticity and is thus a desirable trait.
Here we examine the change in environmental robustness over time during neutral evolution and consider runs with both random and constant ICs.

Given the size of the networks (there are $2^{500} \approx  3 \times 10^{150}$ different initial configurations) a sampling approach was necessary.  To estimate the environmental robustness, a random set of  initial node states was taken, and the output phenotype was measured.  This process was repeated for $1\,000$ different initial conditions (chosen randomly for each dynamics simulation) and the average fraction of initial conditions where the target phenotype was produced for an individual is called $R_d^{ind}$.  The average of $R_d^{ind}$ was taken over all the networks in the population to give $R_d$, a measure of the environmental robustness (the subscript $d$ refers to the dynamic nature of this measure of environmental robustness).  
This quantity was then further averaged over 50 evolutionary runs to reduce statistical error, and measured at the same intervals as in Section \ref{r_m} after maximum fitness is attained.

\begin{figure}[t]
\centering
\includegraphics[width=\linewidth]{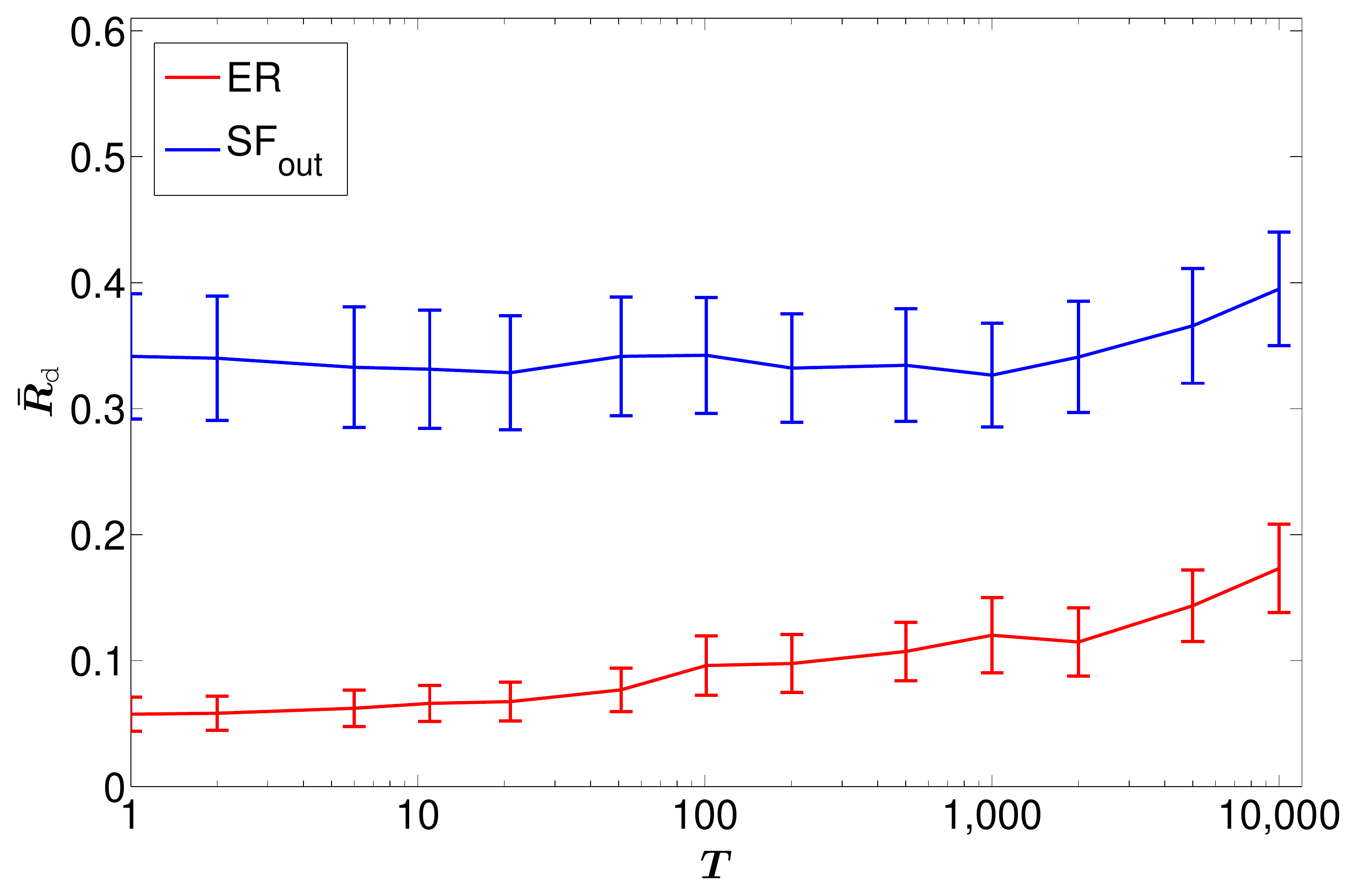}
\caption{Mean environmental robustness $\bar{R}_d$ versus $T$, with constant ICs for $\mathcal{L}=\{10\}$. The SF$_{\rm{out}}$ topology naturally possesses greater environmental robustness. Over time the robustness of SF$_{\rm{out}}$ changes insignificantly. The ER topology, on the other hand, almost doubles its robustness over the period measured.  However, the ER topology still remains much less robust than the SF$_{\rm{out}}$ topology. 
}\label{r_ic1_1x10}
\end{figure}

Results for  the mean environmental robustness $\bar{R}_d$ for  $\mathcal{L}=\{10\}$ targets are shown for random ICs  in Fig. \ref{r_ic0_1x10}  and for runs with constant ICs in Fig. \ref{r_ic1_1x10}.  Firstly we observe that the SF$_{\rm{out}}$ topologies are more environmentally robust than the ER topologies are, an effect that mirrors the mutational robustness. Secondly, the runs with random ICs show a marked increase of $R_d$ during neutral evolution.  The runs with fixed ICs  show a modest increase of $R_d$ during neutral evolution, especially for the ER networks.

When random ICs are used, it may be possible for selection to increase the environmental robustness of the population, for similar reasons to the arguments for the increase in mutational robustness.     During the evolutionary runs those genotypes that are more environmentally robust are more likely to have progeny that generate the required phenotype.  Thus they are more likely to survive than less envrionmentally robust genotypes.     Furthermore, we can make an estimate of the minimal environmental robustness ($R_d^{min}$) for individuals to simply be sustained in the population. Let us assume individuals possess maximal mutational robustness and mutations affect their environmental robustness negligibly. Under these conditions, at least one of the four genotypes must produce the target phenotype again. This gives a value of $R_d^{min}=0.25$.  As the solution must propagate through the population, and is not fully mutationally robust, this is the lower bound and the actual value of $R_d$ required may be significantly larger.

For constant ICs there is no minimum environmental robustness which explains in part why the values of $\bar{R}_d$ are lower than for random ICs.    Moreover, there is no direct  selection pressure for environmental robustness that acts on runs with constant ICs.
We observe that $\bar{R}_d$ does increase for ER networks.  The reasons for this are not clear.  A possible cause could be a correlation between mutational robustness, which is affected by selection, and environmental robustness.  Another possibility would be an entropic effect where diffusion on a neutral space is more likely to find phenotypes with higher robustness.

Although they are not shown, the $\mathcal{L}=\{100 \times 1\}$  targets show similar trends to the $\mathcal{L}=10$ targets:  the SF$_{\rm{out}}$ topology is more environmentally robust than the ER topology and furthermore both topologies show a modest increase in $R_d$ under neutral mutation for both random and fixed ICs.

\section{Discussion}\label{conclusions}

In this paper we have studied the interplay between evolvability and robustness of  Boolean threshold dynamics models with ER and SF topologies.       Our main conclusions are:

{\em 1) The enhanced evolvability of SF networks over ER depends on the phenotype under selection.}

  \noindent We confirm the results of~\citet{cluzel_nat} that networks with SF$_{\rm{in}}$ topology are more evolvable towards single oscillatory targets than ER networks are, and find the same enhanced evolvability for the SF$_{\rm{both}}$ topology and also for the SF$_{\rm{out}}$ topology that most closely resembles the global topology of biological GRNs.   The evolvability advantage of the SF$_{\rm{out}}$ topology over the ER topology increases significantly for multiple targets with the same length.
By contrast, for multiple targets of a different length, the difference in evolvability between SF$_{\rm{out}}$ and ER topologies is reduced, and if the targets are coprime, the differences are even smaller.   We also studied targets where $100$ of the $500$ nodes were given a fixed output.  Here again the differences in evolvabilities between the topologies was greatly reduced compared to the $L=10$ oscillatory outputs studied in~\citet{cluzel_nat}.    These results show that the evolvability advantage of a particular network topology depends strongly on the phenotype under selection.

{\em 2) SF networks show more oscillatory phenotypes and greater synchronicity than ER networks do.}

\noindent We compared the likelihood of finding an oscillatory output of period $L \neq 1$ at a randomly chosen output node in networks without any evolutionary selection. The SF networks show a greater probability of finding oscillatory outputs than the ER networks do.    Similarly, once a population has reached maximum fitness under evolution towards an $L=10$ target for a single output node, a different node in the SF networks is much more likely than an ER network to show  $L=10$ oscillatory output:  SF networks show much greater {\em synchronicity} than ER networks do.    
We argue that the increased evolvability towards oscillatory targets observed for SF networks is due to the fact that such phenotypes are more naturally generated by these topologies.


{\em 3) SF$_{\rm{out}}$ networks are more mutationally robust than ER networks are. For both network topologies, the mutational robustness increases under neutral mutations.}
\noindent One might naively expect that the enhanced evolvability would lead to a reduced mutational robustness.  However, as has been pointed out by numerous authors~\citep{bloom2007evolution,bloom2007neutral,aldana2007,wag_RS,wag_nat_pers,Daniels2008,Draghi2010}, this simple expectation does not necessarily hold for populations.    Even without selection, SF topologies show a greater propensity than ER networks to exhibit
oscillatory phenotypes. This difference suggests that the number of different genotypes that generate the same oscillatory phenotype is larger in SF networks, i.e.\ that the neutral spaces are larger.   Larger neutral spaces may explain how greater mutational robustness can correlate with greater evolvability~\citep{wag_RS}.  However, to firmly establish that the evolvability advantage of the SF networks can be explained by properties of the neutral space requires further investigation.  For example, one must show that larger neutral spaces have evolutionary access to a greater phenotypic diversity.

Another possible factor is the heterogeneity introduced by the SF$_{\rm{out}}$ topology. The measure of mutational robustness treats each node in the network equally (a biologically reasonable assumption), whereas, in an SF$_{\rm{out}}$ topology, nodes are not equal: most nodes trivially have a single outgoing connection, whilst a few have a large number, the so-called `hubs' of the network. The hubs in the SF$_{\rm{out}}$ topology may play an important role, and mutations to these may be most effective, but rare as hubs only comprise a small part of the network. However, when mutations do occur to hubs (or to nodes closely connected to a hub), there is the potential for large scale influence of the mutation.

We also find that the mutational robustness increases under neutral mutations, an effect we attribute to the fact that more mutationally robust genotypes are more likely to have progeny that survive under future mutations~\citep{van_nim}.

{\em 4) SF$_{\rm{out}}$ networks are more environmentally robust to changes in initial conditions than ER networks are, and both networks types exhibit an increase in environmental robustness under neutral mutations.}
\noindent The increased environmental robustness observed in SF topologies may be due to the comparatively large number of nodes acting in cohort to produce the required oscillatory signal -- a feature shown by the observed high synchronicity in SF topologies. This large group of synchronised nodes may act to increase the number of initial states that converge to the required attractor, as perturbations to node states may be ``damped out'' by the overall tendency of the group to produce a particular oscillation. By contrast, the smaller groups of synchronised nodes in ER networks may be more sensitive to the initial states of nodes.


 
For random ICs, the environmental robustness increases under neutral mutations for reasons  similar to those that explain how  mutational robustness increases: more environmentally robust genotypes are more likely to survive in future generations when the environment changes.   We find a similar, but more modest increase in environmental robustness in ER networks for fixed ICs, where there should be no selection for more robust phenotypes.  At present it is not clear what causes this increase, but it may be correlated to the increase in mutational robustness.

An important question is whether the conclusions listed above hold only for the particular models we studied, or whether they are valid more generally.
We show in ~\ref{wag_model} that similar conclusions  also hold for a different threshold model~\citep{wag_1994} that is frequently applied to study GRNs.     Nevertheless, the evolutionary dynamics of the threshold models we study here is considerably simpler than that which occurs in nature and so further work is needed to probe to what extent these conclusions hold for biological GRNs.  At present such studies are very challenging because they require knowledge of the evolutionary history of GRNs, much of which is not easily accessible.       The conclusions from this work that will most likely carry over  are: (a) that the relationship between robustness and evolvability can depend on the class of phenotypes under selection; and  (b) an organism's GRN can become both more evolvable and more robust simply though changing its overall topology. 

\bibliographystyle{elsarticle-harv}
\bibliography{paper2}

\begin{appendix}

\section{Comparison with the Wagner model}\label{wag_model}
\begin{center}
\begin{table}[t]
\caption{The results for evolution towards $L=10$ targets with the A.Wagner model. $\tilde{\tau}$ is shown for both constant and random ICs. ER and SF$_{\rm{out}}$ topologies perform similarly under constant ICs.
However, intrinsic robustness to ICs may play a more important role.}\label{wag_meds}
\vspace{0.05in}
\begin{tabularx}{0.95\linewidth}[b]{| l | X | X |}
\hline
ICs & ER & SF$_{\rm{out}}$\\ \hline
Constant & $25$ & $25$ \\
Random & $19\,000$ & $3\,800$ \\ \hline
Relative & $850$ & $150$ \\
\hline
\end{tabularx}
\end{table}
\end{center}

To examine the generality of the dynamical properties introduced by the SF$_{\rm{out}}$ topology, we made a comparison with another threshold model that was proposed by \citet{wag_1994} and that has been investigated extensively with respect to GRNs. The essential differences between the Wagner model and the one of~\citet{cluzel_nat}, is the use of different node states and update rules. In the original Wagner model, the gene states for node $i$ are either $\sigma_{i}=1, 0, -1$. The update rules differ by using purely threshold rules, with no modified rule for single incoming degrees. The node state $\sigma_{i}=0$, occurs when the input sum $S_i=0$. Node states of $1$ and $-1$ correspond to our ``on'' and ``off'' states considered in the~\citet{cluzel_nat} model.

We investigated dynamics with the essence of the Wagner model. 
We note that nodes with $\sigma_i=0$ make no contribution to the dynamics in the next time step and given a continuous range of weights, such nodes will only occur when lacking any incoming connections. As a result, nodes with $\sigma_i=0$ will always remain in this null state throughout a dynamical run, rendering them ineffective within a network. Effectively, the update rules only differ from the~\citet{cluzel_nat} model because we can abandon the modified rule for single incoming connections. The effect of this rule is now incorporated into the behaviour of all nodes, as when a node is in its ``off'' state, a signal is still affective on the nodes it connects to. This is due to the ``off'' state now corresponding to $\sigma_i=-1$, resulting in a non-zero amount being added in the threshold sum $S_i$, even when the node is ``off''.

To examine the networks using the Wagner model, we firstly consider for what parameters the system enters the chaotic phase. 
 In the~\citet{cluzel_nat} model, networks were within the ordered phase and need the same to be true here for comparison and biological realism. It was found that for the adapted ER topology (or SF$_{\rm{in}}$ topology), a phase transition occurs for an average connectivity of $\langle k \rangle=1$. This gives a critical parameter of $K_c \rightarrow \infty$ for the adapted Poisson distribution. As a consequence, networks with this topology (or SF$_{\rm{in}}$) cannot be produced in the ordered phase. This is an effect due to local noise propagation in the Wagner model being more likely than in the~\citet{cluzel_nat} model (see \ref{ord_cha}).
As a result we considered ER graphs with a standard Poisson distribution
\begin{equation}\label{stan_poi}P_{ER_{std}}(k)=\frac{K^ke^{-K}}{k!}\end{equation}
rather than an adapted one. For the standard Poisson distribution, the phase transition occurs at $K_c=2.075$. This allowed us to compare non-adapted ER networks and SF$_{\rm{out}}$ (SF$_{\rm{out}}$ also possesses a non-adapted ER incoming degree distribution here) networks at the previous average degree connectivity, whilst remaining in the ordered phase.

\subsection{The evolvability of SF networks using the Wagner model is again greater than that of ER networks}
As in Section \ref{sec:1x10}, we evolved networks towards random $L=10$ targets over 50 evolutionary runs. The median generation numbers for a population to reach maximum fitness, $\tilde{\tau}$, are shown in Table \ref{wag_meds} for each topology under random and constant ICs. Evolution with constant ICs is a much simpler task for both, with $\tilde{\tau}$ reduced by more than an order of magnitude for both topologies. Interestingly, the SF$_{\rm{out}}$ topological advantage has vanished in this case. For random ICs, however, the results are dramatically different. The SF$_{\rm{out}}$ topology now outperforms the ER topology by roughly the same factor as in the~\citet{cluzel_nat} model. Both topologies struggle to a much greater extent to adapt under random ICs. This is likely to be a consequence of local noise propagation being greatly increased in this model. 

\subsection{Networks with the Wagner model show more oscillatory behaviour}\label{sync_wag}
\begin{figure}[t]
\centering
\includegraphics[width=\linewidth]{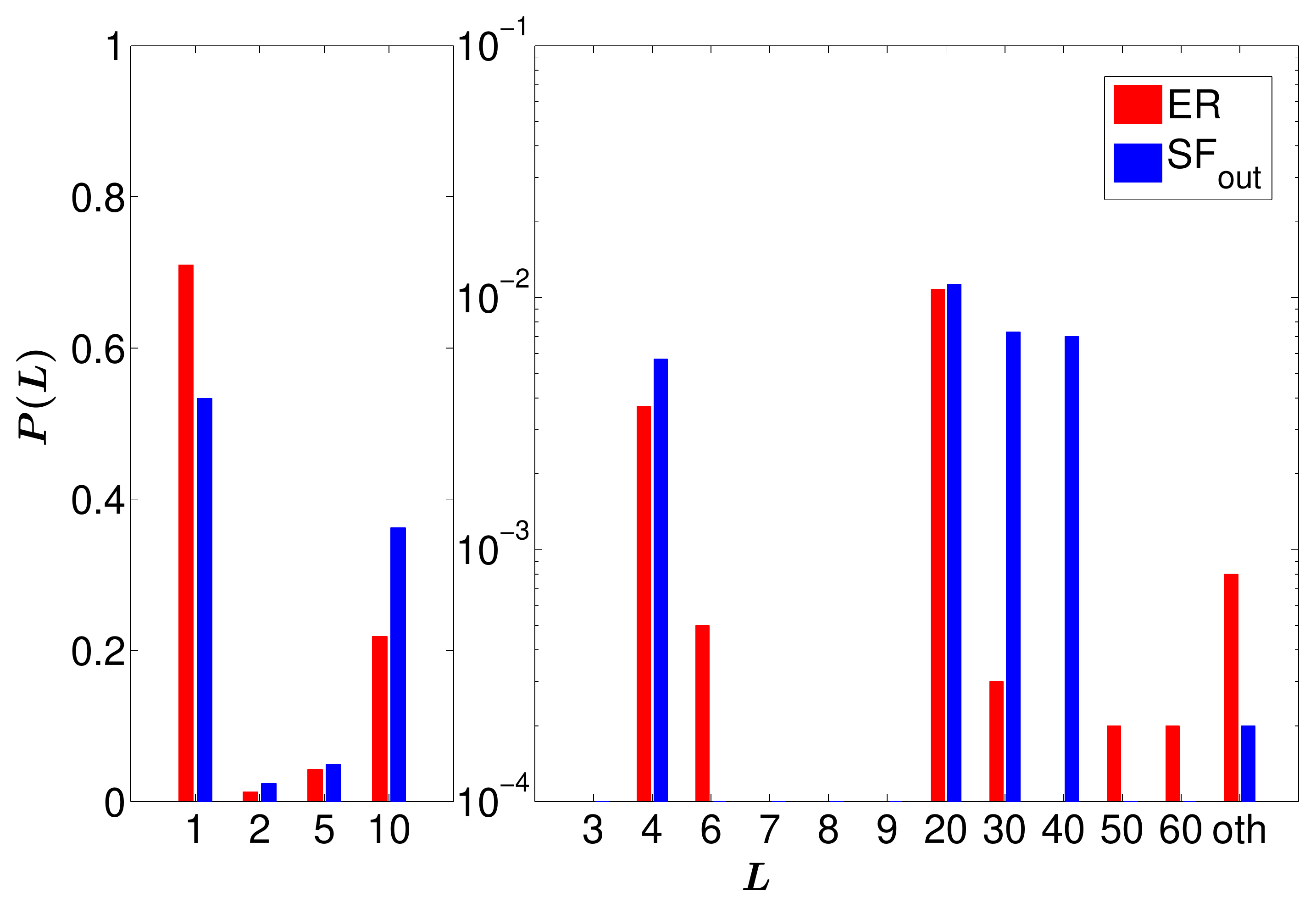}
\caption{The probability a random node has a given period, $L$, in the first generation where maximum fitness is attained, within the Wagner model. The probability is derived from the mean fraction of measured nodes taken over 50 independent evolutionary runs. The ``oth'' column refers to other periods, i.e. all periods not labelled. As in the model of~\citet{cluzel_nat} the SF$_{\rm{out}}$ topology exhibits greater synchronisity.}\label{sync_wag_g}
\end{figure}

As in Section \ref{sf_adv}, the synchronicity in networks under evolution towards a single $L=10$ target was examined but with the Wagner model. The period of each node in the first maximally fit population was measured, with an average taken over 50 evolutionary runs. A comparison of the probability of measuring each period is presented in Fig. \ref{sync_wag_g}. As for the~\citet{cluzel_nat} model, the probability of a node having the same period as the evolved output ($L=10$) is significantly higher in the SF$_{\rm{out}}$ topology than in the ER topology. In the Wagner model, oscillations spread much more widely with both topologies having a much smaller fraction of $L=1$ nodes. This is due to an increase in local noise propagation under the Wagner update rules. 

\subsection{Mutational and environmental robustness of networks with the Wagner model also increase with time after maximal fitness is reached}\label{wag_robust}
\begin{figure}[t]
\centering
\includegraphics[width=\linewidth]{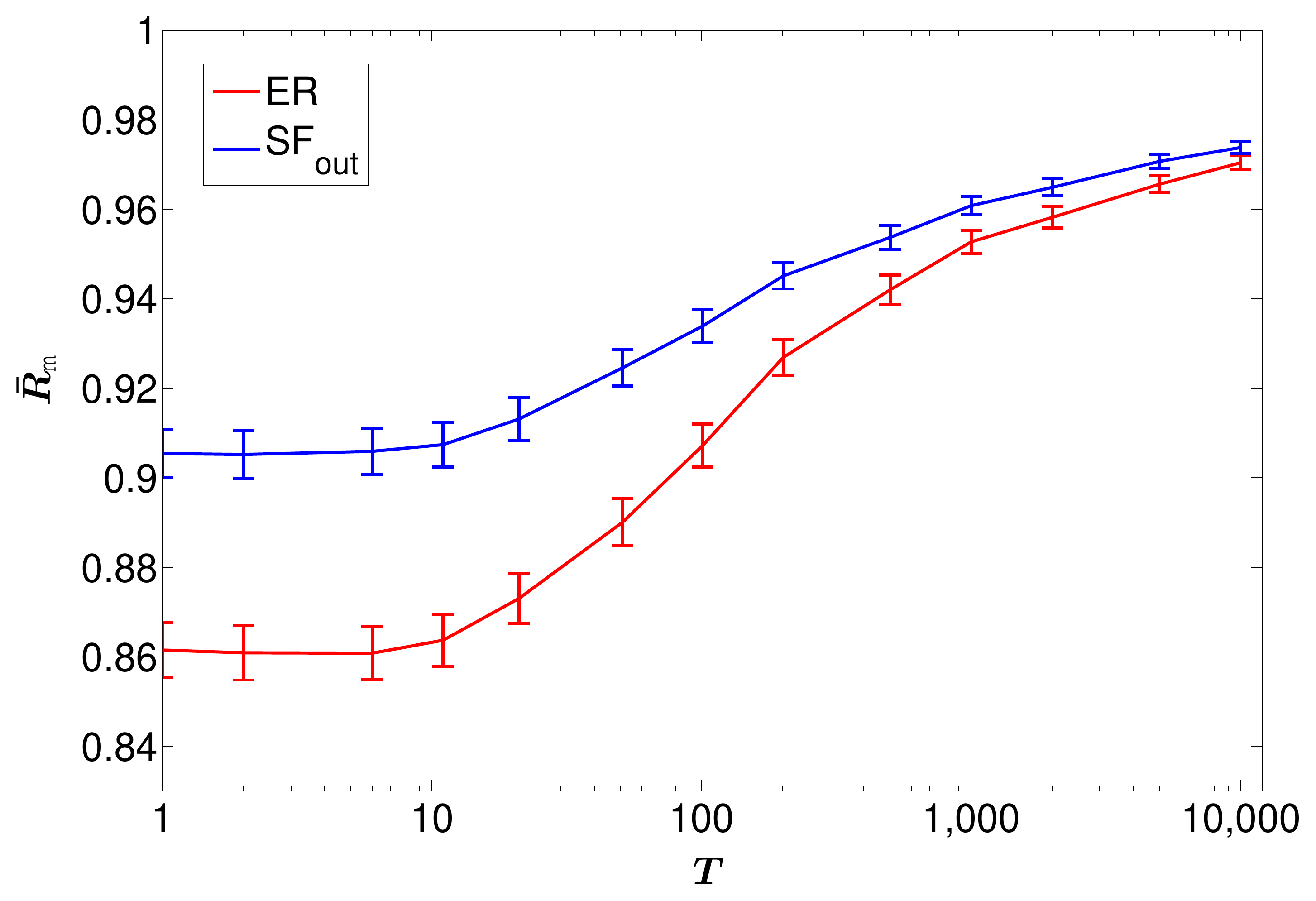}
\caption{Mean mutational robustness $\bar{R}_m$, against $T$ for evolution of $L=10$ targets in the Wagner threshold model. As found for the~\citet{cluzel_nat} model, neutral evolution can optimise muational robustness over time. The relative change in mutational robustness attained in this case is larger than that in the~\citet{cluzel_nat} model, whilst the typical maximal mutational robustness values attained after $10\,000$ generations of neutral evolution are similar.}\label{w_r_m}
\end{figure}

\begin{figure}[!t]
\centering
\includegraphics[width=\linewidth]{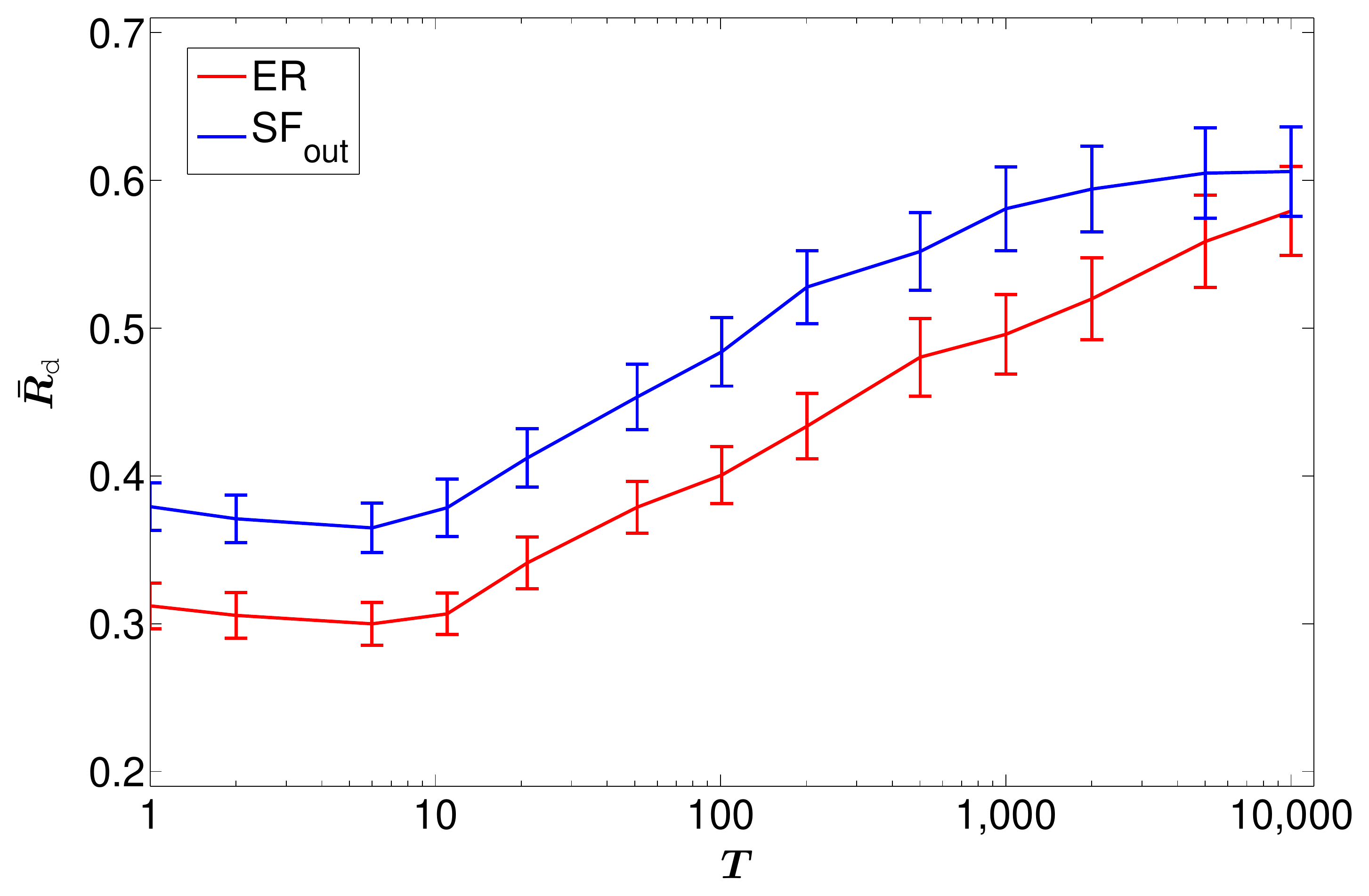}
\caption{Mean environmental robustness $\bar{R}_d$, against $T$ for evolution of $L=10$ targets in the Wagner threshold model for random ICs. Selection increases environmental robustness in both topologies. The limit of this increase appears to be reached towards the end of the neutral evolutionary period, levelling off at around $\bar{R}_d\approx0.6$. The overall lower level of environmental robustness is a consequence of increased local noise propagation in the model.}\label{w_r_ic0}
\end{figure}

In Section \ref{robustness}, we considered the mutational and environmental robustness of networks evolving towards $\mathcal{L}=\{10\}$ targets. These experiments made use of the update rules used by~\citet{cluzel_nat}. In this section, we confirm the overall results presented previously are not a consequence of those update rules, and are a general feature of threshold networks.
The same method of measurements were used as in Section \ref{robustness}.

Fig. \ref{w_r_m} shows the average change in mutational robustness after the population reaches maximal fitness. We find, once again, that SF$_{\rm{out}}$ topologies exhibit greater robustness throughout and that both topologies increase their mutational robustness over time. For the Wagner model the mutational robustness changes to a greater extent for both topologies than is observed for the model of~\citet{cluzel_nat}.  
This difference is likely due to the greater extent of local noise propagation in the Wagner model. 

\begin{figure}[t]
\centering
\includegraphics[width=\linewidth]{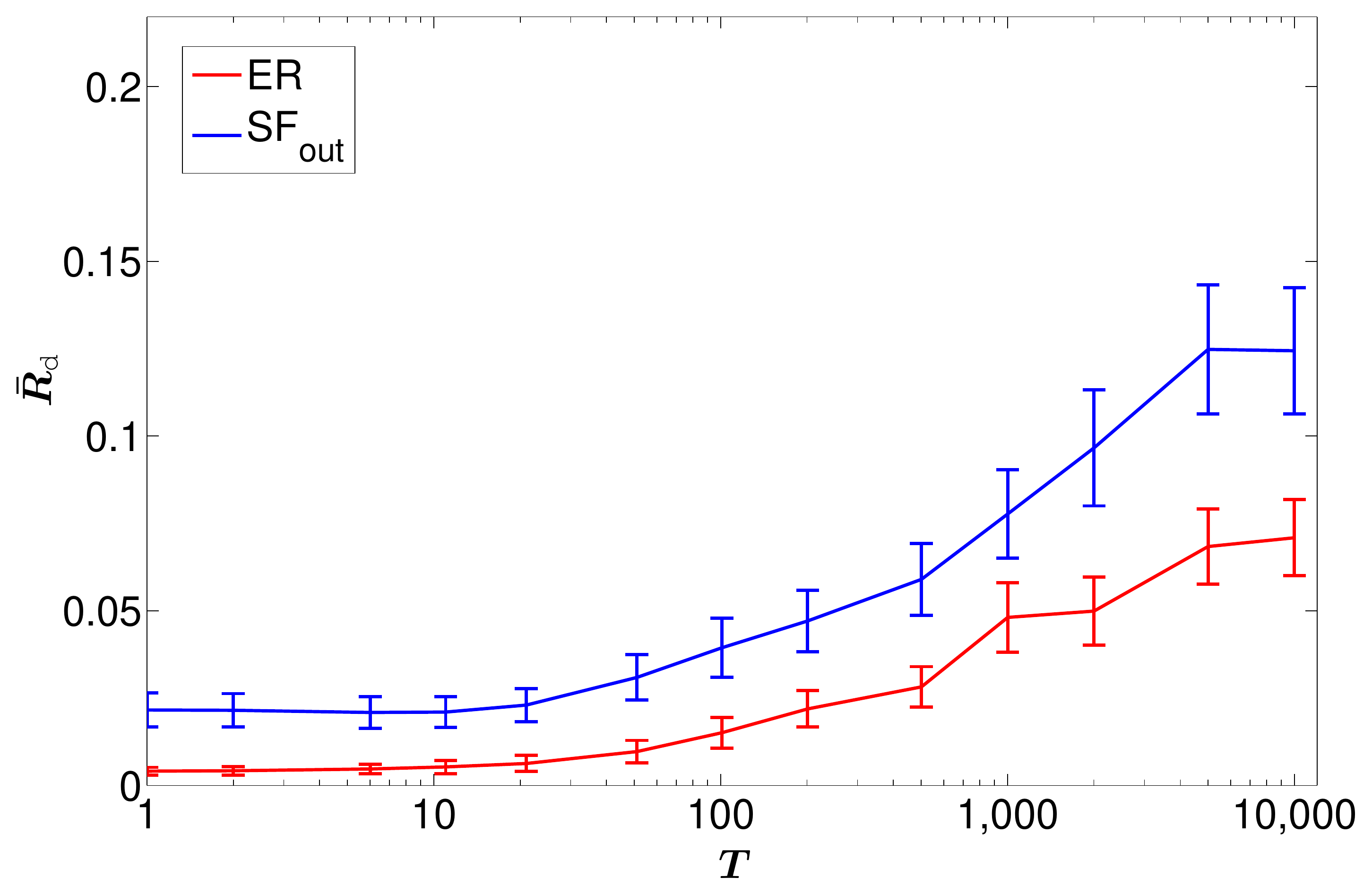}
\caption{Mean environmental robustness $\bar{R}_d$, against $T$ for evolution of $L=10$ targets in the Wagner threshold model for constant ICs. Both topologies demonstrate a significant increase from very low initial levels. The increase generated may be consequence of the selection for mutational robustness also affecting environmental robustness. 
}\label{w_r_ic1}
\end{figure}

Fig. \ref{w_r_ic0} and Fig. \ref{w_r_ic1} show the change in average environmental robustness over the maximally fit evolutionary period, for both random and constant ICs. As before, in the case of random ICs, environmental robustness can be optimised directly by selection. A significant increase in environmental robustness is observed in this case, although for both topologies, the typical maximum environmental robustness attainable starts levelling off around $\bar{R}_d\approx0.6$. This is different to the~\citet{cluzel_nat} model, where a higher environmental robustness was generated by both topologies. Again this difference may be  a consequence of local noise propagation being stronger in the Wagner model. The inability to evolve more complete environmental robustness requires further research. The analysis of the attractor properties within the Wagner model would probably be helpful. The result for constant ICs in Fig. \ref{w_r_ic1} shows that under neutral mutation there is a clear optimisation for both topologies. The networks naturally possess very low environmental robustness at $T=1$, particularly for the ER topology. 

Fundamentally the results are very similar to before with the~\citet{cluzel_nat} model. SF$_{\rm{out}}$ topologies are consistently  more evolvable than  the ER topology, whilst both topologies increase their mutational robustness and environmental robustness under neutral mutation

\section{Calculation of order-chaos phase transition}\label{ord_cha}

\begin{figure}[!t]
\centering
\includegraphics[width=\linewidth]{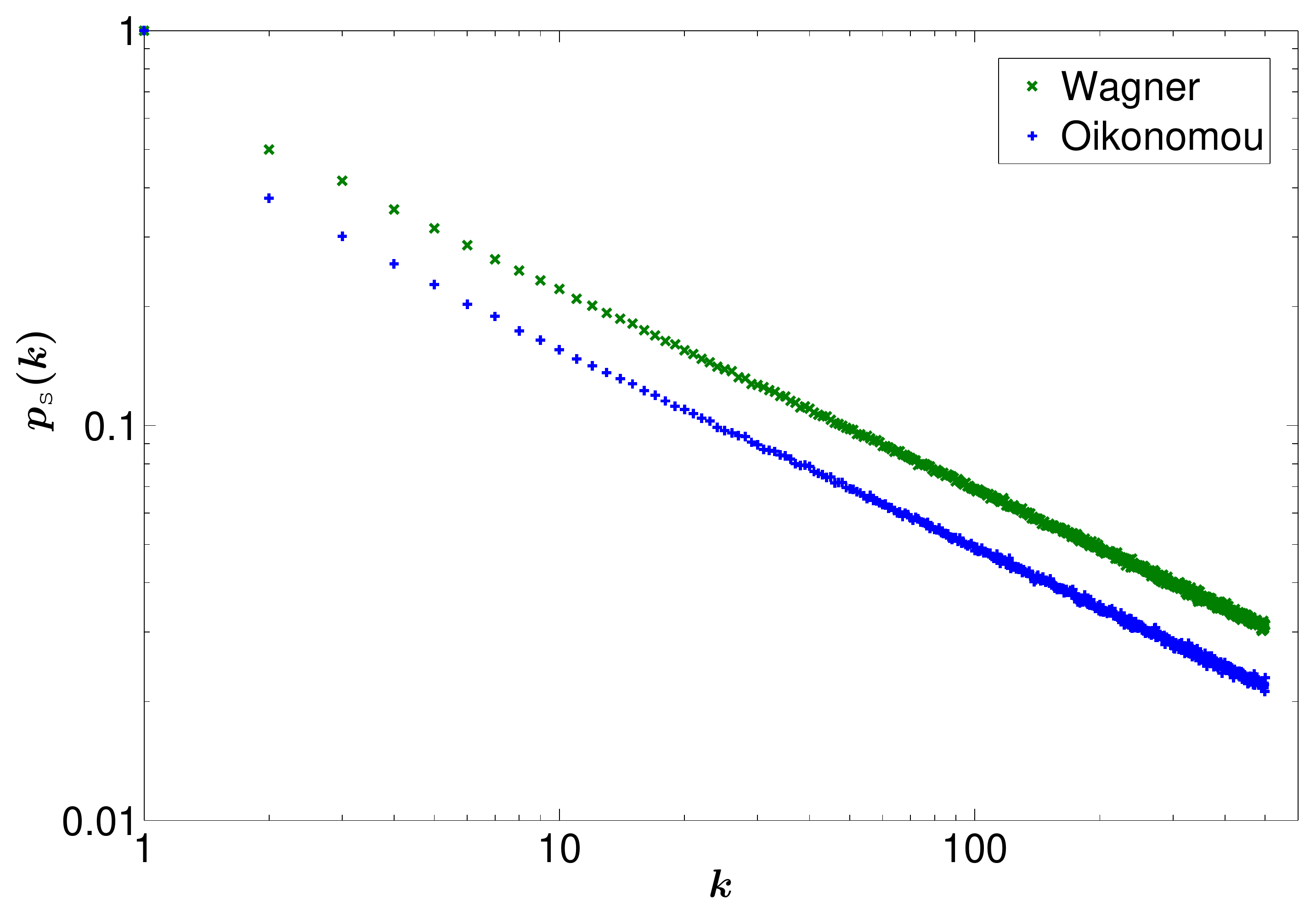}
\caption{The probability of local noise propagation, $p_s(k)$ against $k$ for both the~\citet{cluzel_nat} model and the~\citet{wag_1994} model. Local noise propagation is stronger in the Wagner model. Both distributions are approximately $\propto k^{-0.5}$.}\label{noise_prop}
\end{figure}

In order to calculate the order-chaos transition, we followed the method of~\citet{cluzel_nat}. Let us consider two different sets of ICs, $\{\sigma_i^{(1)}\}$ and $\{\sigma_i^{(2)}\}$. The difference between these sets over time can be measured as with the Hamming distance as
\begin{equation}d(t)=\frac{1}{N}\sum_{i=1}^N|\sigma_i^{(1)}(t)-\sigma_i^{(2)}(t)|\end{equation}
The annealed approximation of~\citet{derrida86} measures whether noise propagation diminishes or extremifies in the limit of the Hamming distance between the two configurations tending to 0
\begin{equation}\label{M_noise}M=\left.\frac{\partial d(t+1)}{\partial d(t)}\right|_{\lim{d(t)\to0}}=\sum_{k_i=1}^\infty P(k_i).k_i.p_s(k_i)\end{equation}
where $M$ measures the noise propagation in the network. When $M>1$, the network is in the chaotic phase, whilst $M<1$ is the ordered phase. For $M=1$, the critical phase is obtained. $P(k_i)$ is the usual degree distribution, whilst $p_s(k_i)$ is the probability of local noise propagation. This measures the probability that if one of the inputs to a node is flipped at time $t$, this will change the output of the node at time $t+1$. For threshold models this is a function of $k_i$, unlike in the Boolean function models. Here we calculate numerically the values of $p_s(k_i)$ for both the update rules in the~\citet{cluzel_nat} model, as well as the Wagner model. These two probability functions are presented in Fig. \ref{noise_prop}. The Wagner model has much greater local noise propagation due to an ``off'' state still passing a signal to a node.

Using these distributions the critical parameter values are calculated for the various topologies using Eq. \eqref{M_noise} and presented in Table \ref{crit_paras}.

\begin{center}
\begin{table}[t]
\caption{The critical parameter values for the phase transition in the various topologies.}\label{crit_paras}
\vspace{0.05in}
\begin{tabularx}{0.95\linewidth}[t]{| l | X  X |}
\hline
Topology & Oikonomou & Wagner\\ \hline
ER$_{adapted}$ & $K_c=3.54$ & $K_c\to0$ \\
SF & $\gamma_c=2.42$ & $\gamma_c\to\infty$ \\
ER$_{standard}$ & $K_c=3.83$ & $K_c=2.08$ \\
\hline
\end{tabularx}
\end{table}
\end{center}

\section{Total number of sequences for each repeat length}\label{number_of_outputs}
There are a finite number of possible configurations a sequence of bits of a given length may
take. Two sequences are regarded as identical if they can be mapped on to each
other through cyclic permutation. If a sequence can be broken down into a
multiple number of shorter sequences, it is invalid. As such, the number of possible
sequences $A(n)$, for a sequence of a length $n$, is given by the following
\begin{equation}A(n) = \frac{2^n-\sum_{k \in \mathcal{F}}kA(k)}{n}\end{equation}
where $\mathcal{F}$ is the set of all factors of $n$ and $A(1) = 2$ (i.e. 0 and 1).

\section{Importance of mutation method with respect to increased robustness}
\begin{figure}[t]
\centering
\includegraphics[width=\linewidth]{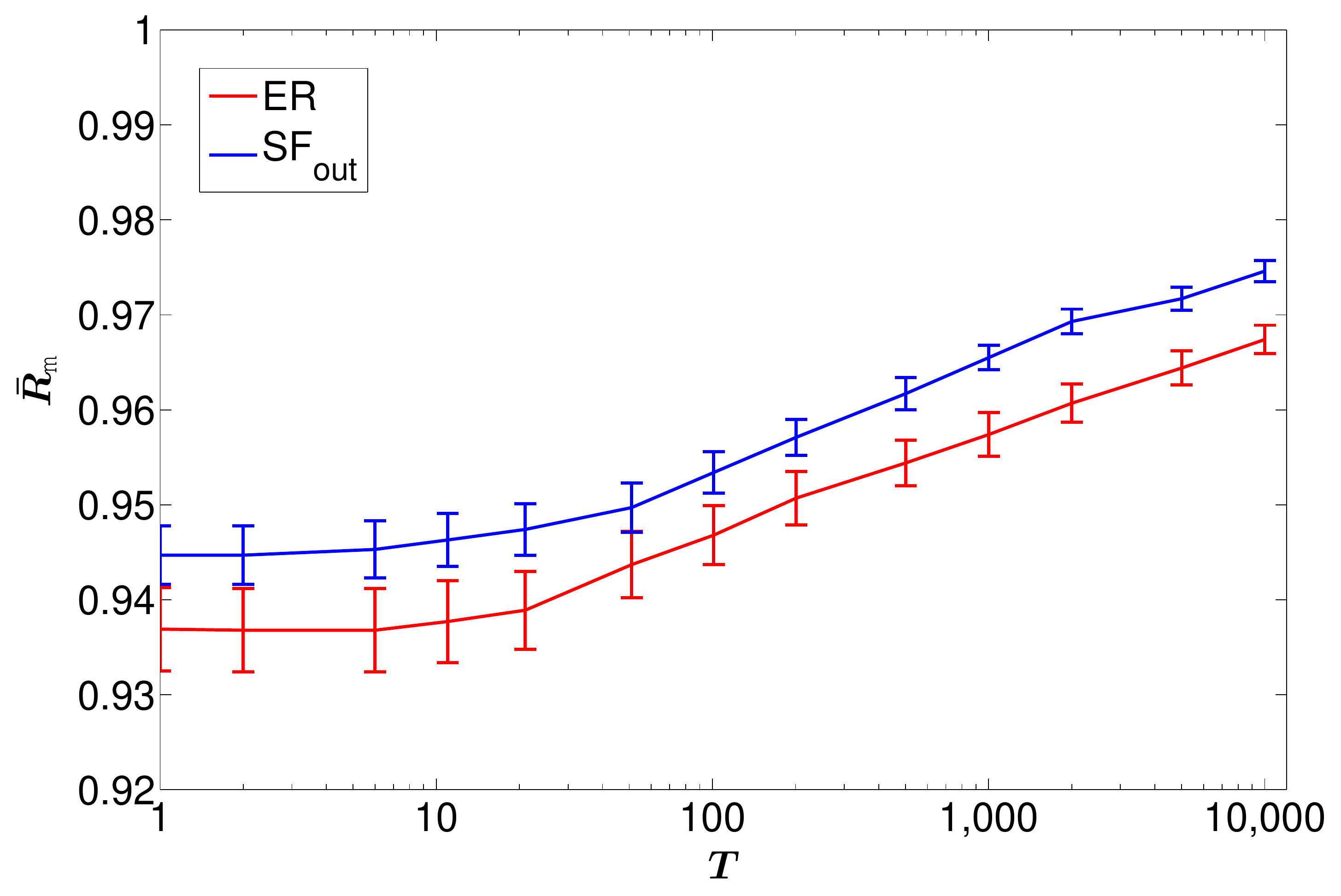}
\caption{Mean mutational robustness $\bar{R}_m$, against $T$ for evolution towards $\mathcal{L}=\{10\}$ with modified weight mutation method. As previously shown, SF$_{\rm{out}}$ topologies possess greater robustness than their ER counterpart indicating that the previous mutation rules do not bias our result.}\label{r_m_1x10_new}
\end{figure}

One possible explanation for the increase in mutational robustness of the SF$_{\rm{out}}$ topology would be a possible bias introduced by the weight mutation rules. This bias could arise in the following way. Given that any edge, incoming or outgoing, can have its weight mutated for the node under mutation, the probability of selecting an incoming node for a hub in the SF$_{\rm{out}}$ topology will be much smaller. Explicitly, there are many more outgoing edges that could be randomly selected for mutation. To assess whether this mutation method was introducing erroneous results, we performed runs towards $\mathcal{L}=\{10\}$ with modified mutation rules. Instead of allowing the weight of any edge touching a given node to be mutated, we only allowed weight mutations of the edges incoming to that node. As both topologies have the same incoming degree distribution, this removes any possible bias.

A graph of the change in average mutational robustness, $\bar{R}_m$, for 50 evolutionary runs is shown in Fig. \ref{r_m_1x10_new}. As can be seen there is limited difference between this and the original, demonstrating any bias does not affect the evolution of mutational robustness in this case.

\section{Population Diversity \label{diversity}}

\begin{figure}[t]
\centering
\includegraphics[width=\linewidth]{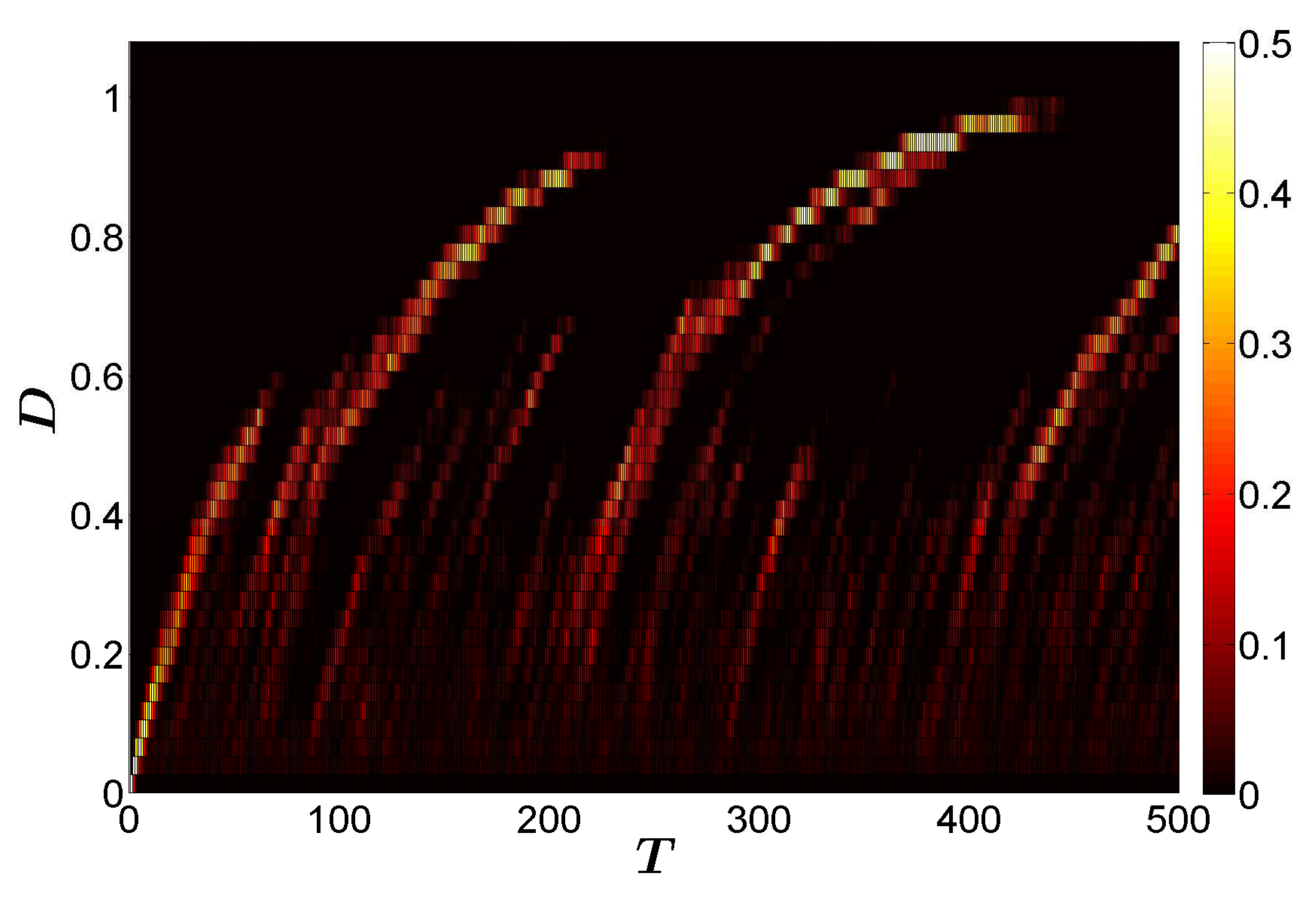}
\caption{The heat of a square represents the fraction of pairs of networks in the population separated by a genetic distance of $D$ at time T. As can be seen, the genetic distance gradually increases as neutral mutations are incurred to each network. The increasing ``hot'' lines demonstrate the population diffusing through genotype space creating a genetically diverse population. The sudden drops in diversity, occassionally observed, correspond to a new most recent common ancestor taking over the population.}\label{heat_map}
\end{figure}
                                                                                                                                                                                     To determine whether the diversity of a  population  genotypes undergoing neutral mutations,  we measured the genetic difference between every pair of inviduals each generation after maximal fitness is reached. The fractional genotypic distance $D(A,B)$, between two networks $A$ and $B$ is given by
\begin{equation}D(A,B)=\frac{1}{2M}\sum_{i=1}^{N}\sum_{j=1}^{N}f(w^{A}_{ij}-w^{B}_{ij})\end{equation}
where
$$f(x)=
\begin{cases} 
1 &\text{if } x\neq0 \\
0 &\text{if } x=0
\end{cases}
$$
and $M$ is equal to the average number of entries in a network's weight matrix, given by $M=N\langle k \rangle$. Values for $D(A,B)$ are taken over all pairs $A,B\in\{population\}$, and then the values are binned in distances of $50/2M$ up to the maximal value of $1$.

These measurements were performed on several different types of evolutionary run. A heat map of these distances is displayed in Fig. \ref{heat_map} for a typical evolutionary run (once maximum fitness is reached) for evolution towards a $\mathcal{L}=\{10\}$ target. This figure demonstrates several features of an evolutionary run. The gradually increasing ``hot'' curves show the population spreading out from each other in genotype space as independent mutations are incurred. The gradient of the ``hot'' line decreases as mutations start occuring on top of each other. However, the ``hot'' lines can still increase to large values (tending to $D\approx1$ for some lines), indicating that most of the mutations are neutral ones and lead to a large population diversity.

An interesting feature of the plot are the occasional dramatic decreases in diversity. These are indicated through the collapse of a ``hot'' line to a much lower value on a new curve, corresponding to the introduction of a new most recent common ancestor for the population. This ancestor may have been selected for or simply fixed through random sampling.

\end{appendix}
\end{document}